\documentclass[twocolumn]{aastex62}

\newcommand{\arcs}{$^{\prime\prime}$} % Arcseconds
\newcommand{\beq}{\begin{equation}\begin{aligned}}
\newcommand{\eeq}{\end{aligned}\end{equation}}

\newcommand{\msun}{M$_\odot$}

\usepackage{scalerel,amssymb}
\usepackage{amsmath}
\usepackage{enumitem}
\usepackage[]{url}
\usepackage{lineno}
\usepackage{CJKutf8}
\usepackage[caption=false]{subfig}
\defcitealias{karachentsev2013}{K13}
\defcitealias{kourkchi2017}{KT17}

\accepted{\today}
\submitjournal{ApJ}

\shorttitle{Isolated Dwarfs in the Local Volume}
\shortauthors{Carlsten et al.}

\begin{document}
\begin{CJK*}{UTF8}{gbsn}

\title{A Sample of Nearby Isolated Dwarf Galaxies: A First Look at the Mass Function of Field Dwarfs}

\correspondingauthor{Scott G. Carlsten}
\email{scarlsten@gmail.com}

\author[0000-0002-5382-2898]{Scott G. Carlsten}
\affil{Department of Astrophysical Sciences, 4 Ivy Lane, Princeton University, Princeton, NJ 08540}

\author[0000-0001-9592-4190]{Jiaxuan Li (李嘉轩)}
\affiliation{Department of Astrophysical Sciences, 4 Ivy Lane, Princeton University, Princeton, NJ 08540}

\author[0000-0002-5612-3427]{Jenny E. Greene}
\affil{Department of Astrophysical Sciences, 4 Ivy Lane, Princeton University, Princeton, NJ 08540}

\author[0000-0001-8251-933X]{Alex Drlica-Wagner}
\affiliation{Fermi National Accelerator Laboratory, P.O.\ Box 500, Batavia, IL 60510, USA}
\affiliation{Kavli Institute for Cosmological Physics, University of Chicago, Chicago, IL 60637, USA}
\affiliation{Department of Astronomy and Astrophysics, University of Chicago, Chicago, IL 60637, USA}

\author[0000-0002-1841-2252]{Shany Danieli}
% \altaffiliation{Hubble Fellow}
% \affil{Department of Astrophysical Sciences, 4 Ivy Lane, Princeton University, Princeton, NJ 08540}
% \affiliation{Department of Astrophysical Sciences, 4 Ivy Lane, Princeton University, Princeton, NJ 08540, USA}
\affiliation{School of Physics and Astronomy, Tel Aviv University, Tel Aviv 69978, Israel}

\begin{abstract}
We present the results of the Exploration of Local VolumE Survey - Field (ELVES-Field), a survey of the dwarf galaxies in the Local Volume (LV; $D<10$ Mpc) over roughly $3,000$ square degrees, focusing on the field dwarf population. Candidates are detected using a semi-automated algorithm tailored for low surface brightness dwarfs. Using tests with injected galaxies, we show the detection is $50\%$ complete to $m_g\sim20$ mag and $M_\star \sim 10^6$ \msun. Candidates are confirmed to be true nearby dwarfs through distance measurements including redshift, tip of the red giant branch, and surface brightness fluctuations. We identify isolated, field dwarfs using various environmental criteria. Over the survey footprint, we detect and confirm 95 LV dwarfs, 44 of which we consider isolated. Using this sample, we infer the field dwarf mass function and find good agreement at the high-mass end with previous redshift surveys and with the predictions of the IllustrisTNG simulation.  This sample of isolated, field dwarfs represents a powerful dataset to investigate aspects of small-scale structure and the effect of environment on dwarf galaxy evolution.

%Additionally, the completeness of this sample allows us to infer the quenched fraction of field dwarfs at this mass for the first time. We find that, at higher dwarf masses ($M_\star \gtrsim 10^7$ \msun), essentially all field dwarfs are star-forming but that at $M_\star \lesssim 10^7$ \msun~ $\sim30\%$ of field dwarfs appear to be quenched. Finally, we find that isolated dwarfs are noticeably smaller ($\sim 25\%$) than satellite dwarfs of the same stellar mass, regardless of quenched status.

\end{abstract}

\section{Introduction}
In order to fully understand dwarf galaxy evolution, it is critical to observe dwarf galaxies across a range of environments. Historically, dwarfs in the vicinity of the Milky Way (MW) and M31 have been the most closely studied \citep[e.g. see][and references therein]{mateo1998, koposov2008, mcconnachie2012, simon2019, drlica2020, ibata2007, ibata2014, mcconnachie2009, mcconnachie2018, bell2011, conn2012, martin2006, martin2016, doliva2025_chapter, savino2023, savino2025,tan2025}. In recent years, significant progress has been made in studying dwarfs that are satellites of other MW-like galaxies \citep[e.g][]{irwin2009, kim2011, sales2013, spencer2014, merritt2014, karachentsev2015, bennet2017, bennet2019, bennet2020, tanaka2017, danieli2017, danieli2020, smercina2018, kondapally2018, park2017, park2019, byun2020, davis2021,davis2024, garling2021, mutlu2021, mutlu2024, kim2022, nashimoto2022, crosby2023, fan2023, muller2024,muller2025, wang2025}. This includes the Exploration of Local VolumE Satellites (ELVES)  \citep{carlsten2022} and the Satellites Around Galactic Analogs (SAGA) Surveys \citep{geha2017, mao2020, mao2024}, which each uniformly mapped the satellite systems of statistical samples of host galaxies analogous to the MW. The primary takeaway from this body of work is that the MW and M31 satellite systems largely appear quite typical for hosts of their mass. In particular, the satellite abundance, luminosity function, colors, and sizes of bright MW satellites are well within the spread found in other, more distant satellite systems. 

This work is currently being extended to survey the satellites of lower-mass hosts (e.g. host galaxies of masses similar to the Large and Small Magellanic Clouds) through projects including ELVES-Dwarf \citep{li2025, li_2025_ddo161}, Identifying Dwarfs of MC Analog GalaxiEs \citep[ID-MAGE;][]{hunter2025}, and Magellanic Analog Dwarf Companions And Stellar Halos \citep[MAD-CASH;][]{madcash, carlin2024}, among others \citep{sand2024, mcnanna2024, zaritsky2024, medoff2025, doliva2025}. Combined with studies of dwarfs in nearby galaxy groups and clusters \citep[e.g.][]{ferrarese2012, ferrarese2016, ferrarese2020, munoz2015, eigenthaler2018, venhola2018, venhola2021, habas2020,  poulain2021, su2021, lamarca2021, marleau2024, marleau2025}, we are beginning to understand the effect that the host halo mass has on satellite dwarfs. A few noteworthy results from these studies include the findings that satellite abundance scales with host mass roughly in accordance with theoretical predictions \citep{danieli2023, li2025}, satellite dwarf sizes are largely independent of parent halo mass \citep{carlsten2021a}, and the abundance of star clusters in dwarfs is greater for satellites that reside in more massive host halos \citep{rsj2019, carlsten2021b, hoyer2023, jones2023, khim2024}.

To complement observations of satellite dwarf galaxies, it is important to also characterize a large sample of isolated, field dwarfs. Currently a few hundred field dwarfs are known in the Local Volume (LV; $D\lesssim 10$ Mpc) with robust distance measures \citep[e.g.][]{karachentsev2004, karachentsev2006, karachentsev2007, karachentsev2013, karachentsev2020, jacobs2009, anand2021}. Many groups are working to expand this number by applying specialized low surface brightness (LSB) detection algorithms to modern deep, wide-field imaging surveys \citep{greco2018, zaritsky2019, tanoglidis2021, zhang2025}. In particular, the extremely wide-field Legacy Surveys \citep{decals, bass1, bass2} are proving a gold mine for LSB dwarf detection efforts. The entire footprint (both the north and south portions) has been algorithmically searched for large dwarf candidates in the Systematically Measuring Ultra-Diffuse Galaxies (SMUDGes) Survey \citep{zaritsky2019, zaritsky2022, zaritsky2023}, and much of it has been visually searched by various authors \citep{karachentsev2022_group_LS, karachentsev2023_field_LS, sand2022}. This usage of the Legacy Surveys can be viewed as a trial run for the even wider and deeper Legacy Survey of Space and Time (LSST, \citealt{lsst}) from the Vera C. Rubin Observatory or the wide field and space-based surveys from Euclid \citep[e.g.][]{hunt2025} or the Roman Space Telescope. The main roadblock to these efforts is the difficulty in getting distances to the many dwarf candidates detected. In the context of searching for satellites of nearby MW analogs in the LV, often a majority of detected LSB dwarf candidates turn out to be background interlopers, and this will be significantly exacerbated when searching for LV field dwarfs. Even with the difficulty of getting distance confirmation, there have been several recent notable detections of individual nearby dwarfs \citep{sand2022, sand2024, jones2023_pavo, jones2024, mcquinn2023_pegw, mcquinn2024_leomk, casey2023, li2024_hedgehog, mutlu2025, fielder2025}. 

However, a large, \textit{volume-limited} sample of LV field dwarfs with well-established completeness limits where \textit{every} detected dwarf candidate gets distance follow-up has proven elusive due to the difficulty of getting distances. The scientific payoff of such a sample is significant, especially when considering the coverage area and depth of LSST. In particular, we highlight two important scientific questions that such a sample could address. First, the stellar mass function (SMF) of field dwarfs below $M_\star \lesssim 10^8$ \msun\; is not well constrained. Second, it is also largely unknown what the quenched fraction of field dwarfs is in this mass range. In this project, we approach the goal of producing such a sample. Paper I (this work), focuses on the stellar mass function. In Paper II (Carlsten et al., accepted), we turn our attention to the star formation properties of isolated dwarfs.

 Redshift surveys, like SDSS and the Galaxy and Mass Assembly Survey \citep[GAMA;][]{driver2011, driver2022, baldry2012}, have constrained the stellar mass function of more massive galaxies. However, the completeness of these surveys to low mass dwarfs, particularly quenched and/or LSB dwarfs (for which acquiring redshifts is very difficult), is unclear. Through abundance matching, the stellar mass function can be used to infer the underlying stellar-to-halo-mass relation \citep[SHMR, e.g.][]{nadler2020, danieli2023, monzon2024, kadofong2025}. The slope, normalization, and scatter of the SHMR at dwarf masses are important observations that can constrain the physics of dark matter, galaxy formation, and reionization \citep{read2019_dm, agertz2020, nadler2021, munshi2021, kim2024}. Recent theoretical work has predicted different SHMRs for satellite dwarfs versus field dwarfs \citep{arora2022, christensen2024} due to biased, early formation of satellite dwarfs in the denser environments around massive hosts. However, \citet{kadofong2025} recently found no significant observational evidence for this down to a dwarf mass of $M_\star\sim10^7$ \msun. Robustly determining the field dwarf SMF and, thus, the SHMR at the low-mass end, will require a volume-limited sample of field dwarfs with well-understood completeness.

Creating such a sample is the main goal of the ELVES-Field survey, an extension of the original ELVES Survey to the field environment. In ELVES-Field, we use the wide-field Legacy Surveys to detect candidate dwarf galaxies using well-motivated photometric selections and a specialized detection algorithm tailored to LSB dwarf galaxies. As part of this, we carefully establish our completeness using injection simulations and comparison with previous searches. Finally, ELVES-Field overcomes the distance roadblock by relying heavily on surface brightness fluctuations \citep[SBF;][]{tonry1988, jerjen_field, jerjen_field2, cantiello2018, sbf_calib, greco2020, cantiello2024} as a way to efficiently get distance estimates to large samples of dwarf galaxies. While \emph{HST} and the precise tip of the red giant branch (TRGB) method it offers is the gold standard for distances in the LV, it is entirely infeasible to acquire TRGB distances to all of the dwarf candidates that will be uncovered in a wide-field (i.e. covering thousands of square degrees) dwarf search\footnote{This will change with either the Euclid or Roman surveys where, due to the space-based resolution, TRGB distances will be available for everything in the survey footprint out to distances of several Mpc.}. Similarly many low-mass field dwarfs might be quenched and without nebular emission lines or HI reservoirs \citep{karunakaran2020, putman2021} and can be quite LSB, making acquiring a redshift extremely difficult. 

Due to these reasons and the improving data quality of wide-field ground based imaging, SBF is seeing increasing use in recent years to measure distances to dwarfs in the LV \citep[e.g.][]{sbf_m101, carlsten2020c, carlsten2020b, kim2022, muller2024, li2024_hedgehog, li2025}. SBF can be done with high-quality ground-based imaging, removing the need for \emph{HST} and, with modern wide-field imagers can get distances for hundreds of dwarfs efficiently. The tradeoff is less precise distances with SBF offering $\sim15\%$ distance errors \citep{sbf_calib, greco2020} versus the $\sim 5\%$ errors of TRGB \citep{beaton2018}. However, the ELVES Survey \citep{carlsten2022} has shown that such distance precision is still more than adequate for LV dwarf research. Dwarf searches with the Rubin Observatory will also likely need to make heavy use of SBF measurements \citep{greco2020}. In this way, ELVES-Field is a test bed for developing ways to detect dwarfs and robustly apply SBF in a scalable fashion in preparation for Rubin.

This paper is structured as follows: in Section \ref{sec:survey_design} we describe the survey design and outline the different sources of data used, in Section \ref{sec:detection} we describe the detection of candidate dwarfs and the completeness tests we perform, in Section \ref{sec:distances} we detail how we measure the distances of the detected dwarfs, in Section \ref{sec:dwarf_properties} we describe the resulting catalog of dwarfs, in Section \ref{sec:abundance} we explore the dwarf abundance and infer the stellar mass function, and finally, in Section \ref{sec:conclusions} we provide an overview of the survey and conclude.

%This paper is structured as follows: in Section \ref{sec:survey_design} we describe the survey design and outline the different sources of data used, in Section \ref{sec:detection} we describe the detection of candidate dwarfs and completeness tests we perform, in Section \ref{sec:distances} we detail how we measure the distance of the detected dwarfs,  in Section \ref{sec:dwarf_properties} we give an overview of the properties of the detected dwarf sample, including the environments of the dwarfs, dwarf structure, and color,  in Section \ref{sec:abundance} we explore the dwarf abundance and infer the stellar mass function,  in Section \ref{sec:quenching} we discuss the star-forming properties of the sample and show the field quenched fraction and, finally, in Section \ref{sec:conclusions} we provide an overview of the survey and conclude.

We use the following conventions throughout this paper: all photometry is in the AB system, solar absolute magnitudes are from \citet{willmer2018}, all photometry is corrected for Galactic extinction using the $E(B-V)$ maps of \citet{sfd} and \citet{sfd2}, and the effective radius, $r_e$, is taken along the major axis.

\section{Survey Design and Data}
\label{sec:survey_design}
ELVES-Field aims to catalog a sample of LV field dwarf galaxies over as much sky area as possible with current observational facilities. Similar to the ELVES Survey, we do this in a two-step process. First, candidate dwarfs are detected in DR10 of the Legacy Surveys\footnote{\url{https://www.legacysurvey.org/dr10/description/}} \citep[LSDR10;][]{decals}. We only make use of the southern portion of the Legacy Surveys ($\delta \lesssim 30^\circ$) that is acquired with the Dark Energy Camera (DECam) on the 4m Blanco telescope in all filters. The northern part \citep{bass1, bass2} is done differently with a mix of telescopes and is generally less deep; thus we do not include it in ELVES-Field. Second, deeper ground-based data is used for SBF distance measurements. The relatively poor seeing ($\gtrsim 1$\arcs) and shallow depth of LSDR10 make it unsuitable for SBF measurements even though it is well-suited for candidate detection \citep{carlsten2022}. Unlike in ELVES where SBF follow-up was often done one-by-one after candidates were found in the Legacy Surveys, in ELVES-Field we only search over areas of the sky already covered by deep ground-based imaging. In particular, we only search for dwarfs in areas of the sky where deep Hyper Suprime-Cam (HSC) imaging exists. The superb seeing and depth of HSC on the 8.2m Subaru telescope make it the ultimate ground-based SBF instrument currently in operation. 

The HSC data come from three sources. First is the third public data release from the HSC Subaru Strategic Program \citep[SSP PDR3;][]{hsc, hsc_pdr3}\footnote{\url{https://hsc-release.mtk.nao.ac.jp/doc/}}. Second is archival HSC data reduced and made available in the HSC Legacy Archive \citep{hscla}\footnote{\url{https://hscla.mtk.nao.ac.jp/doc/}}. This includes all HSC observations taken up through the end of 2016. Finally, we download and reduce raw HSC data from the SMOKA archive\footnote{\url{https://smoka.nao.ac.jp/}}. Raw data from SMOKA is reduced using the HSC pipeline \citep{bosch2018}.

In selecting what HSC data to use, we focus only on $r$ and $i$ band observations as these are the suitable filters for SBF \citep{sbf_calib, carlsten2020b}. Additionally, we only use large contiguous areas and avoid cases where, for example, HSC was being used to study a known LV galaxy. Since we only use the southern component of LSDR10, we also only search for HSC coverage below $\delta \lesssim 30^\circ$. In the end, we use $r$ and $i$ band data from both the SSP and HSCLA sources but only $r$ band raw data from SMOKA. While the depth of the HSC SSP is fairly consistent across the footprint (20 min. integration in $i$ and 10 min. in $r$), the depth of the HSCLA and SMOKA data is more inhomogeneous. Interestingly, essentially all of the SMOKA data that we select appear to have been acquired for solar system minor body searches \citep[e.g.][]{sheppard2019}. For these data, the depth is generally comparable to HSC SSP ($\sim 10$ min. in $r$). Some of these SMOKA data were used in the ELVES and ELVES-Dwarf Surveys, demonstrating their usability for SBF measurements.

The HSC data are typically significantly deeper and have better seeing than LSDR10 data, but we still do the dwarf detection in the Legacy Survey data due to it being more homogeneous. Additionally, much of the HSC footprint is only in a single band (either $r$ or $i$) and the multiband coverage of LSDR10 is important for the dwarf detection (in particular, the use of $g-r$ color), as we discuss in \S\ref{sec:detection}.  

With the HSC data in hand, we then select the corresponding footprint in LSDR10 to actually search over for dwarfs. The Legacy Survey imaging is divided into $0.25^\circ \times 0.25^\circ$ `bricks'. We include all bricks that overlap with any of the HSC coverage that we have. Thus, the LSDR10 footprint is slightly larger than what we actually have HSC coverage for. This includes both an extra border area around the HSC footprints and also the filling in of chip gaps in the HSC data. Since much of the SMOKA data comes from solar system minor body searches, it is often not dithered at all, leaving numerous chip gaps in the coadded images. The survey area that we quote below is the searched area of LSDR10 data, but we account for the smaller coverage of the HSC data by defining an `area reduction factor' that accounts for the amount lost to chip gaps and border areas around the HSC pointings. We calculate this factor from the fraction of detected dwarf candidates in the Legacy Survey data that end up having HSC coverage. We discuss the area reduction in more detail below in \S\ref{sec:abundance}. We find the fraction of the LSDR10 area that has HSC coverage to be $\approx$ 0.83. Although we do the dwarf detection in the LSDR10 footprint, we only include dwarfs in the final sample that fall into the HSC footprint as well.

As for the LSDR10 data, we download the data directly from the Legacy Survey server\footnote{\url{https://portal.nersc.gov/cfs/cosmo/data/legacysurvey/dr10/south/}}. For each brick that we include in the search, we download the $grz$ coadded images, the inverse variance maps, the bitmasks, and the \texttt{tractor}\footnote{\url{https://github.com/dstndstn/tractor}} source catalog. 

Our goal in ELVES-Field is to find isolated LV dwarfs. We discuss different definitions of `isolated' in \S\ref{sec:dwarf_properties} and, thus, want to be agnostic to environment at this point in the survey definition. Therefore, we do not mask around any massive LV hosts. We note that several ELVES hosts are in the HSC footprint. We do, however, mask out very massive nearby ($D<30$ Mpc) groups. This includes the Virgo cluster and NGC 5846 group \citep{mahdavi2005}. Both of these are significant mass concentrations with total stellar mass $\gtrsim10^{11.5}$ \msun\,\citep{kourkchi2017} and would contribute substantial numbers of dwarf candidates. Masking them out causes only a small loss in footprint area but simplifies the distance confirmation step and reduces the number of dwarf candidates that we need to process. For each of these two groups, we 
mask out to a projected radius corresponding to twice the estimated virial radius from \citet{kourkchi2017}.

\begin{figure*}
\includegraphics[width=\textwidth]{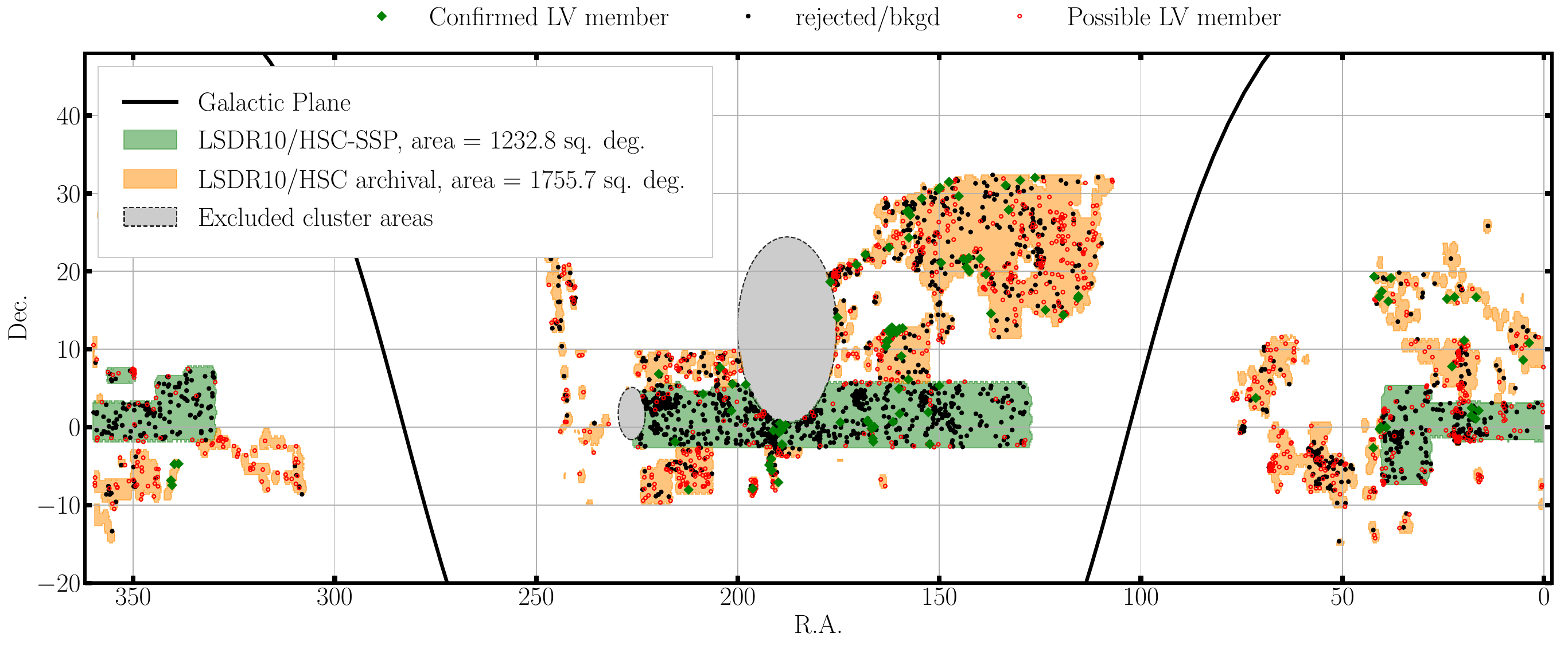}
\caption{The area surveyed for LV field dwarfs in ELVES-Field. The area shown is specifically the Legacy Survey footprint that is used in the object detection. These footprints are chosen because they also have existing deeper archival HSC imaging which is used for the SBF measurements. HSC data coming from the HSC-SSP versus other archival sources are shown in different colors. The two large, gray excluded areas correspond to the Virgo Cluster and the NGC 5846 group and are not searched for dwarfs. The points show the actual dwarf candidates that we detect with the different markers indicating the results of the distance confirmation step in \S\ref{sec:distances}.}
\label{fig:survey_area}
\end{figure*}

The area used for the survey is shown in Figure \ref{fig:survey_area}. In particular, this figure shows the LSDR10 bricks that we include in the search. The bricks corresponding to different HSC data sources are shown in different colors. About $1,230$ square degrees corresponds to HSC-SSP coverage while roughly $1,760$ square degrees comes from archival HSC sources. Out of the archival data, $\sim950$ square degrees comes from the data we reduce from SMOKA while the remaining $\sim810$ square degrees is from the HSCLA. Note that these quoted areas are for the LSDR10 coverage and, thus, are slightly more than the actual HSC coverage as we discuss above. For comparison, the $\sim3,000$ square degrees included in ELVES-Field is several times larger than the $\sim400$ total square degrees included across all hosts in the ELVES Survey.

\section{Detection Algorithm}
\label{sec:detection}
In recent years, many authors have applied semi-automated LSB detection algorithms to wide-field imaging surveys \citep[e.g.][]{bennet2017, greco2018, zaritsky2019, tanoglidis2021, carlsten2020a, li2023a, li2023b, li2025, hunter2025}. While the details differ between algorithms, the detection generally consists of steps to mask higher surface brightness and/or compact background galaxies and then filtering to bring out faint and LSB features. In ELVES-Field we use a similar detection algorithm to that used in the ELVES Survey with some key differences to make the algorithm more scalable for the larger area surveyed here. In the following subsections, we provide detail on the detection algorithm used and our tests to quantify the completeness of the algorithm.

\subsection{Overview of Detection Algorithm}
\label{sec:algorithm}
The detection of candidate LV dwarfs occurs in a three-stage process. First, we make use of the Sienna Galaxy Atlas \citep[SGA;][]{moustakas2023} for large and bright galaxies. Second, for intermediate brightness candidates, we make use of the LSDR10-provided \texttt{tractor} source catalogs. Finally, for very faint and/or LSB candidates, we blur the images and detect LSB sources in a bespoke object detection step. We describe each of these steps along with various masking steps in more detail throughout this section. Figure \ref{fig:det} shows this process for a portion of an individual LSDR10 brick. For each of the $\sim48,000$ bricks that we process, the main steps of the detection pipeline are as follows:

\begin{figure*}
\includegraphics[width=\textwidth]{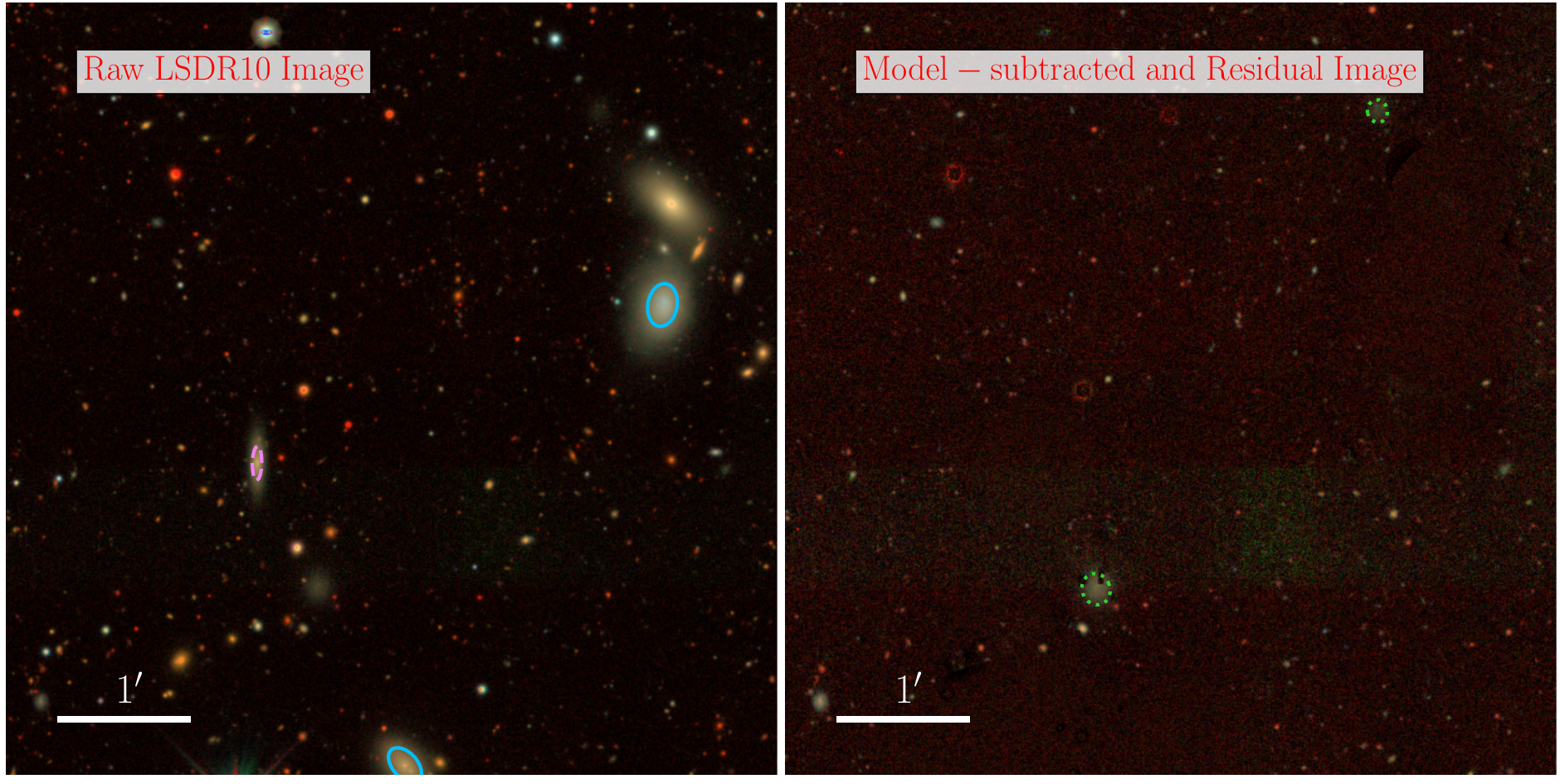}
\caption{A demonstration of the dwarf detection process for a portion of the LSDR10 brick `1849p062'. The left panel shows a $grz$ composite of the original LSDR10 imaging. Detections coming from the Sienna Galaxy Atlas \citep{moustakas2023} and \texttt{tractor} that pass our photometric cuts are shown in blue solid and pink dashed ellipses, respectively. The right panel shows the image after the \texttt{tractor} models for stars, high surface brightness sources, and very red sources are subtracted out. The background in the right panel appears less flat due to a more aggressive image stretch to bring out faint and low surface brightness objects. On this cleaned image, a separate low surface brightness-focused detection step is run. The resulting detections that pass our photometric cuts are shown in the green dotted ellipses.   }
\label{fig:det}
\end{figure*}

\begin{enumerate}
    \item We take the SGA sources that fall within the brick and then mask out their footprint using the LSDR10 bitmask image. The SGA galaxies are saved for analysis later in the pipeline. Additionally, at this stage we mask all other pixels flagged by the bitmask (including e.g. cosmic rays and saturated pixels). Each masked pixel is replaced by noise using the LSDR10-provided inverse variance image. There are an average of $1.2$ SGA entries per brick in our footprint.

    \item{We use the \texttt{tractor} source catalog to subtract out and mask stars, high surface brightness (HSB) sources, and very red sources. Details on the \texttt{tractor} source detection and extraction can be found in the Legacy Survey documentation\footnote{\url{https://www.legacysurvey.org/dr10/description/}}}, but we provide some relevant details here. In brief, \texttt{tractor} simultaneously fits either a PSF, exponential, de Vaucouleurs, or S\'{e}rsic profile to all sources in a brick. This allows for substantially better photometry, especially for crowded areas, than simple source extraction algorithms, e.g. \texttt{SExtractor}, can do. Which model is fit to which source is determined by the $\chi^2$ of the fit. However, \texttt{tractor} will likely over-deblend (`shred') many LSB sources. Still it is trustworthy for stars and HSB sources. Empirically, we find that sources with either $m_g<19$ mag or $\mu_{g, \mathrm{eff}} < 22$ mag arcsec$^{-2}$, where $\mu_{g, \mathrm{eff}}$ is the $g$-band surface brightness averaged within the effective radius, are generally modeled correctly in \texttt{tractor}. These HSB sources from \texttt{tractor} very well might be genuine LV dwarfs, and so we save this list of sources and pass it through the pipeline, along with the SGA sources from above. This contributes an average of $94.9$ detections per brick. Very red objects which here we take to be all things with $g-r > 0.9$ mag are almost certainly not LV dwarfs \citep[][and see below for more detail on the photometric cuts we apply]{geha2017, greco2018, tanoglidis2021} and thus can be disregarded and subtracted out at this stage. We use the \texttt{tractor} fit parameters to generate models of each source that gets subtracted out. Subtracting out, as opposed to simply masking the sources, helps remove extended wings and allows us to recover faint, LSB candidates near to brighter objects (see below for the LSB detection stage). With that said, we also mask these areas to cover up possible residuals from the subtraction. This residual image is shown in the right panel of Figure \ref{fig:det}.

    \item{The $g$ band residual image from the previous step is blurred with a $2$\arcs\;Gaussian filter to bring out faint extended emission. Then we use a \texttt{SExtractor} detection step with a threshold of $3\sigma$ and minimum area of 1000 pixels to detect candidates. These candidates are further required to have \texttt{SExtractor}-measured effective radii $>3.5$\arcs (170 pc at a distance of 10 Mpc). Note that this quoted size is after the convolution with the $2$\arcs\; kernel. This is a conservative cut and less restrictive than the final cut we apply on size (see \S\ref{sec:photo_cuts} below) measured with more accurate S\'{e}rsic fitting on the unblurred image. We save this list of sources and, combined with the SGA and HSB \texttt{tractor} sources, it is passed to the next step of the pipeline. This detection step generates an average of $8.4$ detections per brick.}

    \item{The list of candidate sources found in the three steps detailed above are each fit with a S\'{e}rsic profile to get robust and uniform photometry before applying the photometric cuts. All of the detections from our third, LSB-focused detection step get fit with a S\'{e}rsic profile but only the SGA and \texttt{tractor} detections that already pass the photometric cuts (using either the provided SGA or \texttt{tractor} photometry) get refit in our pipeline. This is simply to reduce the amount of fitting that is needed, particularly since there are many \texttt{tractor} detections per brick. On average, we fit $8.9$ sources per brick. The actual fitting is done using \texttt{pysersic}\footnote{\url{https://github.com/pysersic/pysersic}} \citep{pysersic}. All three bands ($grz$) are fit simultaneously. This fitting stage is fully automated and uses the masks derived from \texttt{tractor} as described above along with the inverse variance images as provided by LSDR10. For the sake of speed, we use \texttt{pysersic} to return the maximum a posteriori S\'{e}rsic parameter estimates instead of sampling the entire posterior distributions. }

    \item{Photometric cuts are applied to the candidates to isolate those that could feasibly be genuine LV dwarf galaxies. We go into more detail below in \S\ref{sec:photo_cuts} where we explain how we derived these cuts. }

    \item{The resulting candidates are now visually inspected to remove false positives, including tidal features of massive background galaxies, Galactic cirrus, faint background spirals, and various image artifacts. In total $\sim36,000$ candidates are visually inspected. The most common false positive is Galactic cirrus constituting $\sim40\%$ of the detections. About $8\%$ of detections pass the visual inspection. This roughly 11:1 false positive rate is a factor of several better than what was achieved in the ELVES Survey due to more robust S\'{e}rsic rather than \texttt{SExtractor} photometry in the ELVES detection pipeline. }

    \item{The S\'{e}rsic fits of the candidates that pass the above inspection are now visually inspected to make sure the automated fit is acceptable. Fits that are clearly problematic based on visual appearance are refit manually. Most commonly, the problems in the fits are due to the automated \texttt{tractor}-based masks being either too aggressive or not covering enough. We refit $\sim25\%$ of the fits at this stage. The candidates that are refit get passed through the photometric cuts again with their updated photometry to make sure they still pass.}

    \item{Finally, candidates are required to have coverage in the HSC imaging data. As mentioned in \S\ref{sec:survey_design}, much of the HSC data is non-dithered with chip gaps present in the final coadded images in addition to significant gaps between pointings. Due to the importance of the HSC data in confirming or rejecting detected candidates as real LV dwarfs, detected dwarfs that fall into chip gaps or outside the FOV of a HSC pointing are not included in further analysis. The fraction of dwarfs detected in the LSDR10 footprint that have HSC coverage is $83\%$. For completeness, in Appendix \ref{app:photometry}, we list the locations of dwarf candidates detected in LSDR10 that are not included in the survey because they had no HSC coverage.}

\end{enumerate}

At this point, the resulting list of candidates is passed on to the distance determination step of the pipeline which we describe in detail in \S\ref{sec:distances}.

\begin{figure*}
\includegraphics[width=\textwidth]{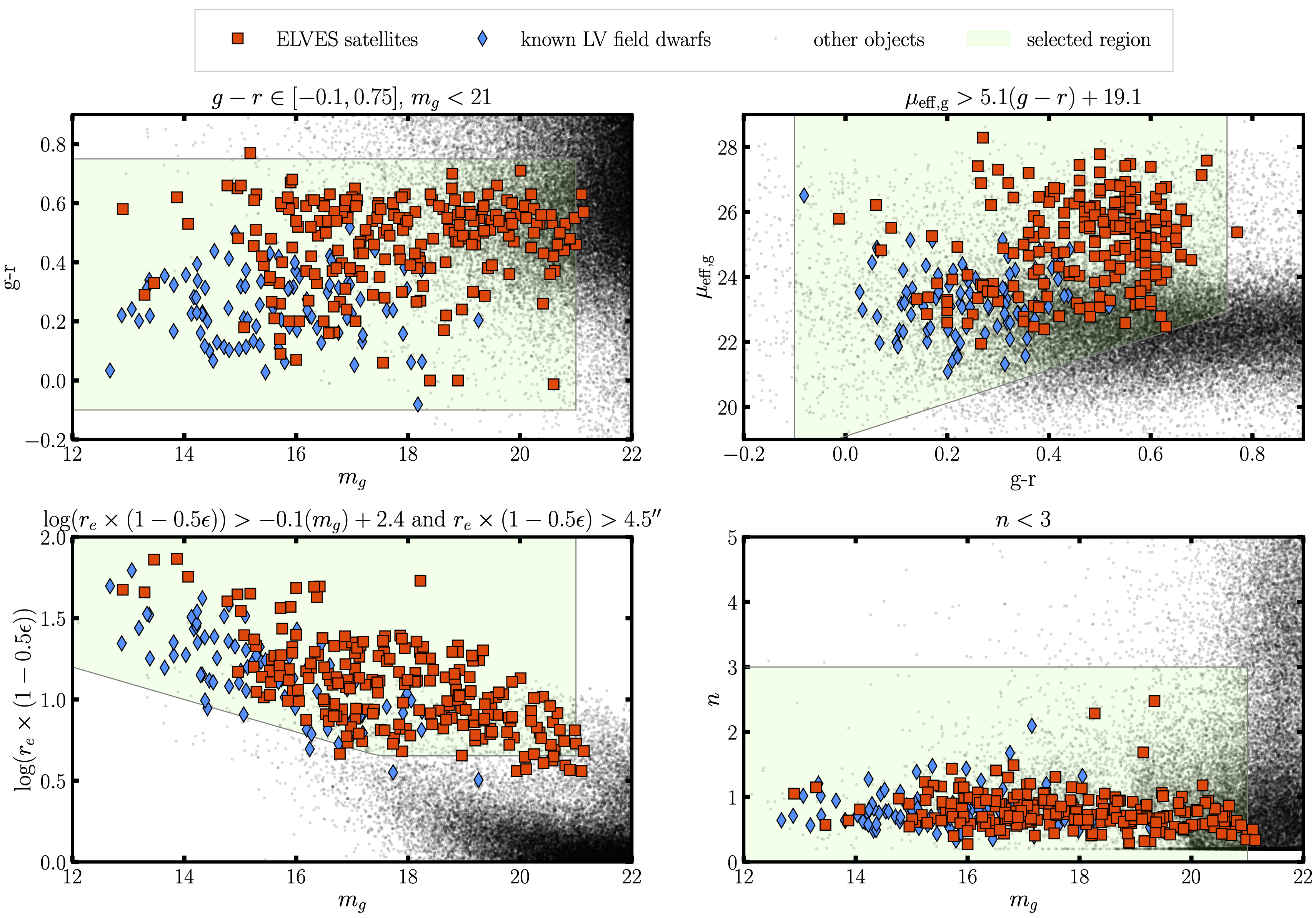}
\caption{An overview of the photometric cuts applied to detected objects to isolate possible LV dwarfs. Each panel shows a different projection of photometric space and the region used to select dwarf candidates. The titles of each panel give the actual selection cut used. Effective radii are measured in arcseconds. In the bottom left panel, $\epsilon$ refers to the ellipticity of the source. For reference, we show known LV dwarfs, including ELVES satellites from \citep{carlsten2022} and field LV dwarfs from \citet{karachentsev2013}. The small black points show the distribution of all sources from Legacy Survey photometry for a representative sample of LSDR10 bricks. }
\label{fig:photo_cuts}
\end{figure*}

\subsection{Photometric Selection Cuts}
\label{sec:photo_cuts}
In order to eliminate a great number of false positives and isolate candidates that can feasibly be real LV dwarfs, we apply stringent photometric cuts on the catalog of sources detected by our detection algorithm. To derive these cuts, we assemble a reference sample of known LV dwarfs with existing S\'{e}rsic photometry. This sample includes two main sources: 1) ELVES satellites from \citet{carlsten2022} and 2) field dwarfs from the Local Volume Galaxy Catalog \citep[LVGC;][]{karachentsev2013}\footnote{Here and throughout the paper, we use a version of the catalog downloaded in April 2024, unless otherwise stated.} with photometry from \citet{carlsten2021a}. The latter sample consists of essentially all known LV dwarfs that are not ELVES satellites that have TRGB distances and that fall in the Legacy Survey Data Release 9 footprint. This earlier version of the Legacy Surveys was the latest when \citet{carlsten2021a} derived S\'{e}rsic photometry for LVGC objects. We apply a cut of $M_g > -15$ mag to these samples to only consider dwarf galaxies. 

This reference sample is shown in Figure \ref{fig:photo_cuts}. Using this sample, we derive cuts in several photometric parameters to eliminate as many background galaxies as possible while retaining most of the known LV dwarfs. Five of these cuts are shown in Figure \ref{fig:photo_cuts}. The cuts we perform are as follows:

\begin{itemize}
    \item{ \textbf{Magnitude:} We select only candidates with $m_g < 21$. This corresponds to $M_g = -9$ mag for a galaxy at the edge of the LV at $D=10$ Mpc. Through our work with the Legacy Surveys \citep[e.g.][]{carlsten2022}, we have found that this is about the limit of detection as lower luminosity (and, hence, lower mass) galaxies will often be too low surface brightness to be detected in the imaging data of the Legacy Surveys. Additionally, fainter galaxies will likely be very small ($r_e \lesssim 3$\arcs~ at $D\sim10$ Mpc) and suffer from significantly more contamination from background galaxies. Thus, it is not fruitful to push to fainter luminosities than this in ground-based imaging surveys.  }

    \item{ \textbf{Color}: Generally, LV dwarf galaxies will be significantly bluer than the majority of background galaxies. Thus we require $g-r \in [-0.1, 0.75]$ mag. As Figure \ref{fig:photo_cuts} shows, even the primarily early-type ELVES satellites are almost all bluer than $g-r=0.75$ mag. Similarly, the LSB galaxy searches of \citet{greco2018} and \citet{tanoglidis2021} find almost all LSB dwarfs to be bluer than $g-r=0.75$ mag. The lower bound in color is used to remove artifacts. Not shown in Figure \ref{fig:photo_cuts}, we also require $r-z \in [-0.1, 0.5]$ mag. These cuts are, again, primarily to just remove artifacts.}

    \item{ \textbf{Size}: The most selective cut that we apply is a cut in size. We apply this cut on the circularized effective radius (i.e. average of the semi-major and semi-minor axes, $r_e \times (1-0.5\epsilon)$, where $\epsilon$ is the ellipticity) measured in arcseconds. We do this to penalize highly elliptical systems as those are often edge-on background disks. We apply both a magnitude dependent cut on size: $\log(r_e \times (1-0.5\epsilon)) > -0.1 m_g + 2.4$ and a size floor: $r_e \times (1-0.5\epsilon) > 4.5$\arcs. As seen in Figure \ref{fig:photo_cuts}, this is the cut that removes the largest number of true LV members in the reference sample. However, we find that relaxing this cut greatly increases the number of false positives. We account for the loss of true LV dwarfs (i.e. false negatives) when quantifying the completeness of the detection algorithm in \S\ref{sec:completeness}. }

    \item{\textbf{Surface Brightness}: Dwarfs are generally low surface brightness and, thus, we apply a color-dependent cut on effective surface brightness of: $\mu_{\mathrm{eff}} > 5.1 ( g-r) + 19.1$ mag arcsec$^{-2}$. The color dependence accounts for the fact that redder dwarfs are more likely to be quenched and thus have lower surface brightness.    }

    \item{\textbf{Concentration}: We apply a simple cut in S\'{e}rsic index: $n < 3$. Dwarfs are found to generally have S\'{e}rsic index $\sim0.7$ \citep{carlsten2021a}, and this can distinguish them from the often more concentrated massive background galaxies.}

    \item{\textbf{Color Gradient}: We find that a significant number of false positives found by the detection algorithm are small, faint background spirals. To help remove these, we apply a cut on the color gradient exhibited by candidates since these background spirals are often redder in the center than the outskirts. The color gradient, $\Delta_{g-r}$, is calculated as the difference in $g-r$ color measured within $0.25\times r_e$ and that measured in an elliptical annulus extending from $0.5\times r_e$ to $r_e$. We then require this gradient to be $\Delta_{g-r} \in [-0.75, 0.1]$. The upper bound was determined through experimentation with a test sample of visually identified background spirals. We chose the threshold so that it removed many of the background spirals while not removing any of the reference sample of known LV dwarfs. The lower bound is to remove some artifacts. We found that this gradient was not robustly measured for many very LSB detections, and so we only apply it for candidates with $\mu_{\mathrm{eff}} < 24$ mag arcsec$^{-2}$.   }

\end{itemize}

As shown in Figure \ref{fig:photo_cuts}, several of these cuts remove genuine LV dwarfs. The magnitude cut, $m_g < 21$ mag, removes a handful of known LV dwarfs, primarily faint, red ELVES satellites. Out of the reference sample, the cuts shown in the Figure cut $5\%$ of the reference galaxies with $m_g < 21$ mag. This does not include the $r-z$ and $\Delta_{g-r}$ cuts, since we do not have $z$ photometry or color gradient measures for the ELVES or LVGC samples. However, we quantify how much these cuts affect completeness in \S\ref{sec:completeness}.

\begin{figure*}
\includegraphics[width=\textwidth]{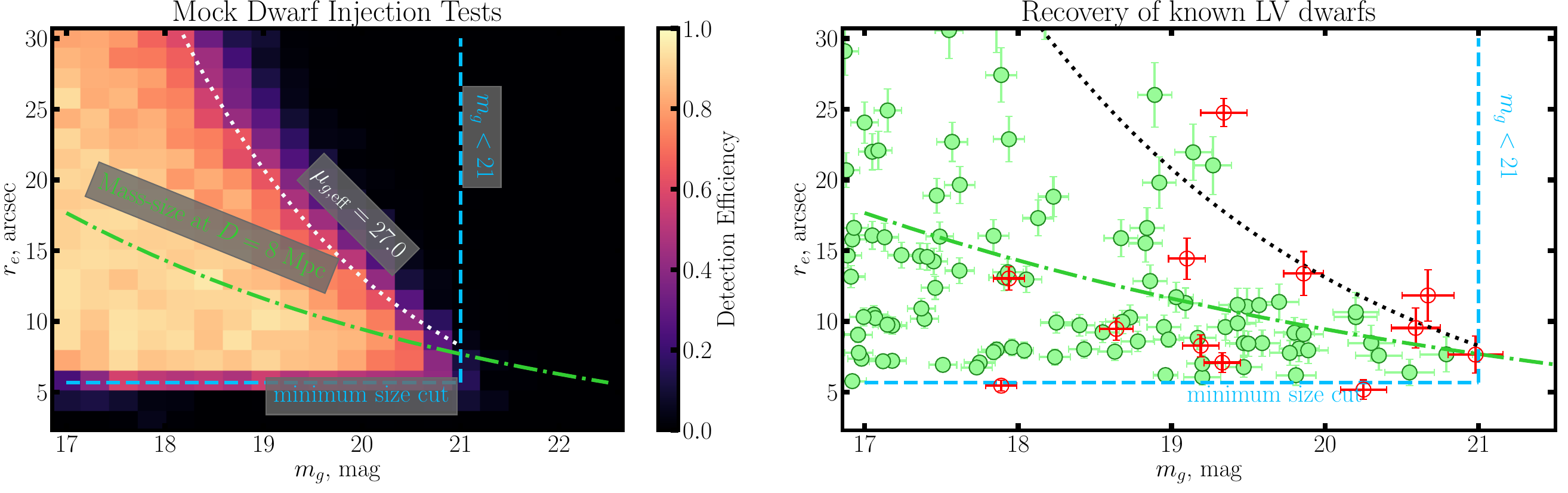}
\caption{The completeness of the survey as shown through injection of artificial dwarfs (left) and recovery of known LV dwarfs (right). For the injection tests, we inject artificial dwarfs across a range in size and magnitude and quantify the recovery fraction. The left panel shows that the recovery is generally $\gtrsim 90\%$ above the size and luminosity cuts applied and brighter than $\mu_{\mathrm{eff}}<27$ mag arcsec$^{-2}$ in surface brightness. The dot-dashed green line shows the mass-size relation of \citet{carlsten2021a} at a distance of 8 Mpc for reference. The right panel confirms this completeness level for actual known LV dwarfs. Filled green circles are LV dwarfs that are recovered by the detection algorithm while empty red circles are missed.}
\label{fig:completeness}
\end{figure*}

\subsection{Completeness Tests}
\label{sec:completeness}
We establish the completeness of our dwarf detection algorithm by injecting artificial dwarfs into the LSDR10 imaging and measuring the fraction of dwarfs recovered as a function of luminosity and size of the injected dwarfs. Since we, in part, use the \texttt{tractor} detection results from LSDR10, to fully test the completeness of the detection, we would need to rerun \texttt{tractor} on the images after we inject artificial dwarfs. However, this is a significant computational challenge that is beyond the scope of this paper. Thus, in testing the dwarf recovery, we use the original LSDR10 \texttt{tractor} results and essentially just rerun our specialized LSB detection step (step 3 in \S\ref{sec:algorithm}). The faintest and lowest-surface brightness detections all come from that third step, so this is a valid test of our completeness towards faint and LSB candidates. However, it is approximate for brighter candidates that come from the \texttt{tractor} results, which might be suffering from shredding. Thus, to supplement the injection tests, we also check the completeness of the algorithm towards known LV dwarfs.

For the injection tests, we randomly select three thousand bricks across the surveyed footprint to inject artificial dwarfs onto. We inject directly into the coadded LSDR10 images. Artificial dwarfs are generated across a grid of apparent $g$ band magnitude and effective radius. Ellipticities and S\'{e}rsic indices are randomly drawn from the observed distributions of these quantities from the ELVES Survey \citep{carlsten2021a}. We consider both cases where the artificial dwarfs are similar to observed LV star-forming dwarfs with $g-r=0.26$ and similar to observed quenched dwarfs with $g-r=0.55$ as these are the average late-type and early-type colors from ELVES, respectively. In the end, we do not find significant difference between the two cases and the results shown here are the average. The actual images of the artificial dwarfs are generated in the DECam filter system using \texttt{ArtPop} \citep{artpop} at a distance of 8 Mpc. The artificial dwarfs are dimmed according to the expected amount of Galactic extinction at the location they are injected. 

After injection, we rerun the detection algorithm and quantify the recovery fraction of the simulated dwarfs. The left panel of Figure \ref{fig:completeness} shows the results of this test. Note that the recovery fraction is shown as a function of the injected (i.e. true) size and magnitude of the artificial dwarfs and is calculated after the photometric cuts of \S\ref{sec:photo_cuts} are applied. This panel shows that the completeness is $\gtrsim 90\%$ for injected dwarfs above the size cut and below the $g$ band magnitude cut down to a limiting surface brightness of $\mu_{\mathrm{eff}}\sim27$ mag arcsec$^{-2}$. This surface brightness is about the same as the $\mu_{0,V}=26.5$ mag arcsec$^{-2}$ limit quoted in the ELVES Survey as both surveys use the same depth Legacy Survey data. In the Figure, we also plot the mass-size relation from \citet{carlsten2021a} assuming a distance of 8 Mpc and intermediate color of $g-r=0.45$. The `minimum size cut' line in the Figure is at $r_e\sim5.7$\arcs\; (not simply 4.5\arcs) as it assumes the average ellipticity of the simulated dwarfs of $0.42$. For reference, the LSDR10 bitmask files we use mask an average of $6\%$ of the pixels. Thus the peak completeness of $\sim 90\%$ is about as expected when considering the area lost due to stars, artifacts, and interference from background galaxies.

We do not incorporate a visual inspection stage in the injection tests but assume that any true LV dwarf would pass the visual inspection. This assumption is based on the fact that all of the $\sim60$ known LV dwarfs in the search footprint pass the visual inspection and, thus, the false negative rate of the inspection must be quite low.

To supplement the injection tests, we also test the recovery of known LV dwarfs. For this, we create a reference sample of all known LV dwarfs in the entire LSDR10-South footprint (i.e. not just in the HSC footprint). This consists of all dwarfs from the LVGC with TRGB distances and all confirmed ELVES satellites. In making this sample, we apply a $M_g > -17$ mag cut to focus only on LV `dwarfs'. We take reference photometry where available from \citet{carlsten2021a} and \citet{carlsten2022}. Only dwarfs with existing reference photometry are included in this test. We apply the photometric cuts from \S\ref{sec:photo_cuts} except for the $r-z$ and color gradient cuts since those are not possible with the available photometry. The resulting recovery of this sample is shown in the right panel of Figure \ref{fig:completeness}. Filled green points are LV dwarfs that we recover with the new algorithm while empty red points are not recovered. There are 203 galaxies in this reference sample and 189 ($93\%$) are recovered. The unrecovered dwarfs are missed for a variety of reasons, including falling in the LSDR10 starmask, the color gradient cut, and being cut by one of the photometry cuts when the photometry is measured by the current algorithm. The photometry of this sample as reported by the current algorithm agrees well with the reference photometry from \citet{carlsten2021a} but there is scatter and some of the dwarfs that pass the cuts in the reference photometry scatter out when remeasured here. The high recovery fraction of known LV dwarfs is encouraging and demonstrates that shredding is not a significant problem in the use of the \texttt{tractor} catalogs.

\begin{figure*}
\includegraphics[width=\textwidth]{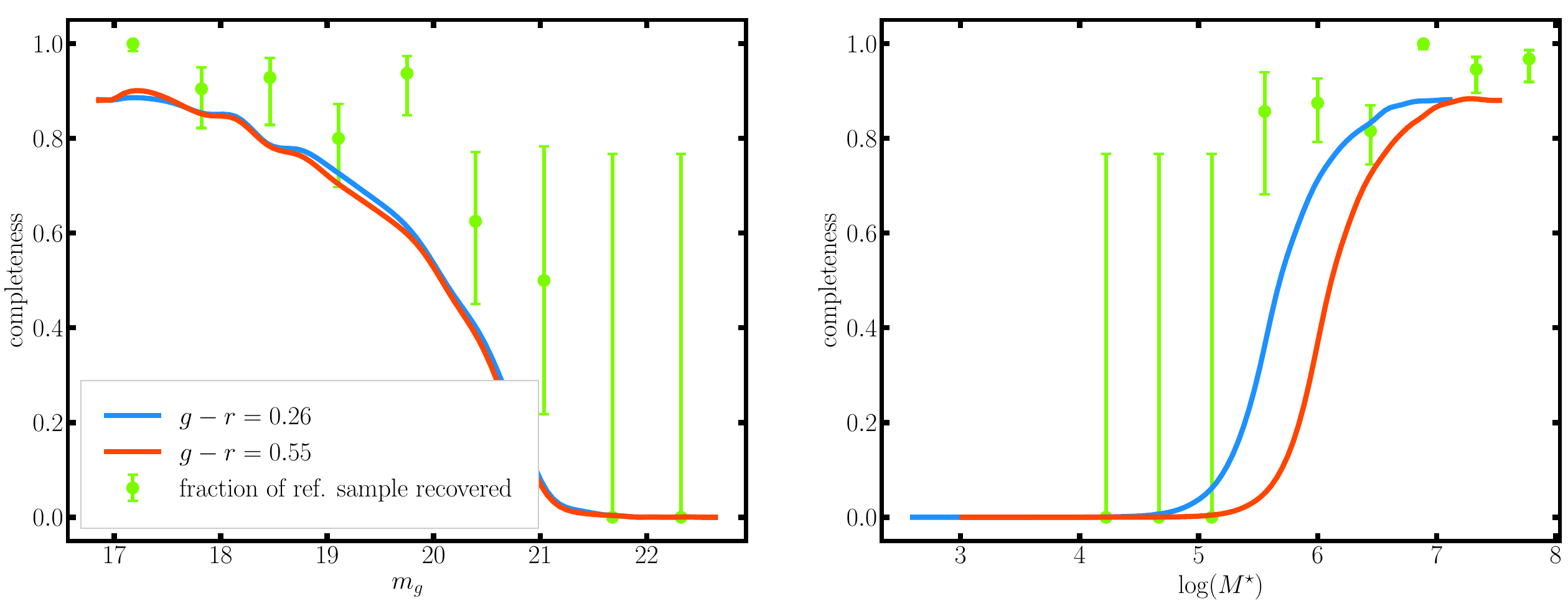}
\caption{Assuming dwarfs are uniformly distributed in space and follow the mass-size relation of \citet{carlsten2021a}, we collapse the completeness estimate shown in Figure \ref{fig:completeness} to an estimate of the completeness as just a function of apparent $g$ band magnitude (left) and stellar mass (right). We do this separately for dwarfs with $g-r$ color characteristic of a star-forming dwarf (blue) and that characteristic of a quenched dwarf (red). The green points show the completeness of the detection algorithm to the same reference sample of known LV dwarfs shown in Figure \ref{fig:completeness}.}
\label{fig:completeness2}
\end{figure*}

In addition to the completeness as a function of magnitude and size, it is useful to have an estimate of the completeness as a function only of stellar mass. To make this estimate, an assumption about the distance and size distributions of LV dwarfs is required. We do this process for both a star-forming-like population of dwarfs with $g-r=0.26$ and a quenched-like population with $g-r=0.55$. We start with a grid of stellar masses, and then, for each of 10,000 iterations, we randomly sample a distance in the range $[0,10]$ Mpc assuming that dwarfs are uniformly distributed in space. We convert the stellar masses to apparent $g$ band magnitudes using the color, distance, and stellar mass conversion given in \citet{delosreyes2024}. We assign angular effective radii using the distance and the mass-size relation from \citet{carlsten2021a} including an intrinsic scatter of 0.18 dex. Then, we estimate the completeness of the detection algorithm by interpolating Figure \ref{fig:completeness} to the appropriate $g$ band magnitude and size. The average completeness curves as both functions of apparent $g$ band magnitude and stellar mass are shown in Figure \ref{fig:completeness2}. Like in Figure \ref{fig:completeness}, the completeness is shown after the photometric cuts of \S\ref{sec:photo_cuts} are applied. In this Figure, we also show the detection completeness of the reference sample of known LV dwarfs from Figure \ref{fig:completeness} which agrees with the injection test results. The green points show the average recovery fraction of the reference sample in each $m_g$ bin and the errorbars show the Bayesian ($1\sigma$) confidence interval, treating the completeness as a binomial proportion parameter with an uninformative Jeffreys prior. Note that for the reference sample, we use the stellar masses reported by \citet{carlsten2021a} which used the color-mass-to-light ratios of \citet{into2013} which will be slightly higher ($\sim0.1-0.2$ dex) than the \citet{delosreyes2024} stellar masses, as shown by the latter authors. We find that the completeness starts to drop fainter than $m_g \sim 19$ mag because some of the dwarfs fail the size cut due to the scatter in the mass-size relation. In the right panel of Figure \ref{fig:completeness2}, we find slightly different expected completeness levels for red dwarfs than for blue dwarfs, with our completeness extending to lower stellar masses for blue dwarfs than red. This is due to a lower mass-to-light ratio expected for a bluer stellar population such that a certain apparent magnitude (e.g. $m_g=20$ mag) corresponds to a lower stellar mass for a blue system than a red one. For red dwarfs, we expect to be complete to down to $M_\star\sim10^6$ \msun\; at the $50\%$ level. We note that this is slightly worse than the completeness claimed in ELVES ($M_\star\sim10^{5.5}$ \msun) due to the higher detection threshold used here and the larger size cut.

\section{Distances}
\label{sec:distances}
The candidate dwarfs cataloged by the detection pipeline must be confirmed via a distance measurement to be genuine LV members. Across the survey footprint, we find $\sim2300$ candidate dwarfs that pass the photometric cuts and visual inspection stage and that have HSC coverage. As we show with the results of this section, the vast majority of these turn out to be background contaminants. Even with the stringent photometric cuts we apply to the detections, only a few percent of the detections turn out to be real LV galaxies. In ELVES-Field, the distance measurements come from three sources: 1) literature TRGB, 2) literature redshift, and 3) new SBF measurements. In each of the following three subsections, we describe how we use each metric in more detail. The next two subsections outline how we combine the different distance measurements to categorize each candidate dwarf as within or outside of the LV and what we do in the cases where we have no conclusive distance constraint for a candidate. The final subsection serves as a summary of how we incorporate distance information for each candidate.

\subsection{Tip of the Red Giant Branch}
\label{sec:trgb}
Of the distance metrics, using TRGB results is the most straightforward. We simply take TRGB distances, where available, from the LVGC \citep{karachentsev2013}. We assume a typical distance error of $5\%$ for TRGB distances \citep[e.g.][]{beaton2018}. In the end, only 27 ($\sim1\%$) of the candidate detections have existing TRGB distances spread fairly uniformly in magnitude across $12 < m_g < 20$ mag.

\subsection{Redshift}
\label{sec:redshift}
While redshift is a critical extragalactic distance tool at large distances, it can be fraught in the LV due to the contribution of peculiar velocities. Peculiar velocities of several hundred km/s due to nearby mass concentrations are common, which can change the distance inferred from Hubble's law by several Mpc. To overcome this, it is common to use `flow models' when inferring distance from redshift that take into account at least some of the nearby mass concentrations, like the Virgo Cluster, \citep[e.g.][]{mould2000}. In this work, we use the redshift-distance relation from \citet{kourkchi2020} which is based on the gravitational trajectory models of \citet{shaya2017}. In this latter work, the Cosmic-Flows3 \citep{tully2016} galaxy catalog is used to calibrate the masses of various nearby mass concentrations, and the orbit trajectories of many nearby galaxies and/or groups are reconstructed numerically. \citet{kourkchi2020} then use this reconstructed velocity field and interpolate it to provide an estimated distance given a redshift and coordinates. The tool is accessed via web\footnote{\url{https://edd.ifa.hawaii.edu/NAMcalculator/}} or API. We note that this orbit reconstruction does not apply to virialized regions around nearby massive hosts and thus these distance estimates are likely spurious for satellites of nearby massive hosts. Since our focus is on field dwarfs, this is not a significant concern for ELVES-Field. Additionally, many satellites of nearby hosts already have SBF distances from the ELVES Survey,

After the visual inspection step, we query SIMBAD\footnote{\url{https://simbad.cds.unistra.fr/simbad/}} \citep{Wenger_2000} for each object to find any existing redshift information. The redshifts are then fed into the \citet{kourkchi2020} distance calculator to provide an estimate of the distance. When the calculator predicts three possible distances for a given redshift, we simply take the median. This is largely only an issue in the infall regions surrounding massive clusters which we already mask out (see Figure \ref{fig:survey_area}).

\begin{figure}
\includegraphics[width=0.48\textwidth]{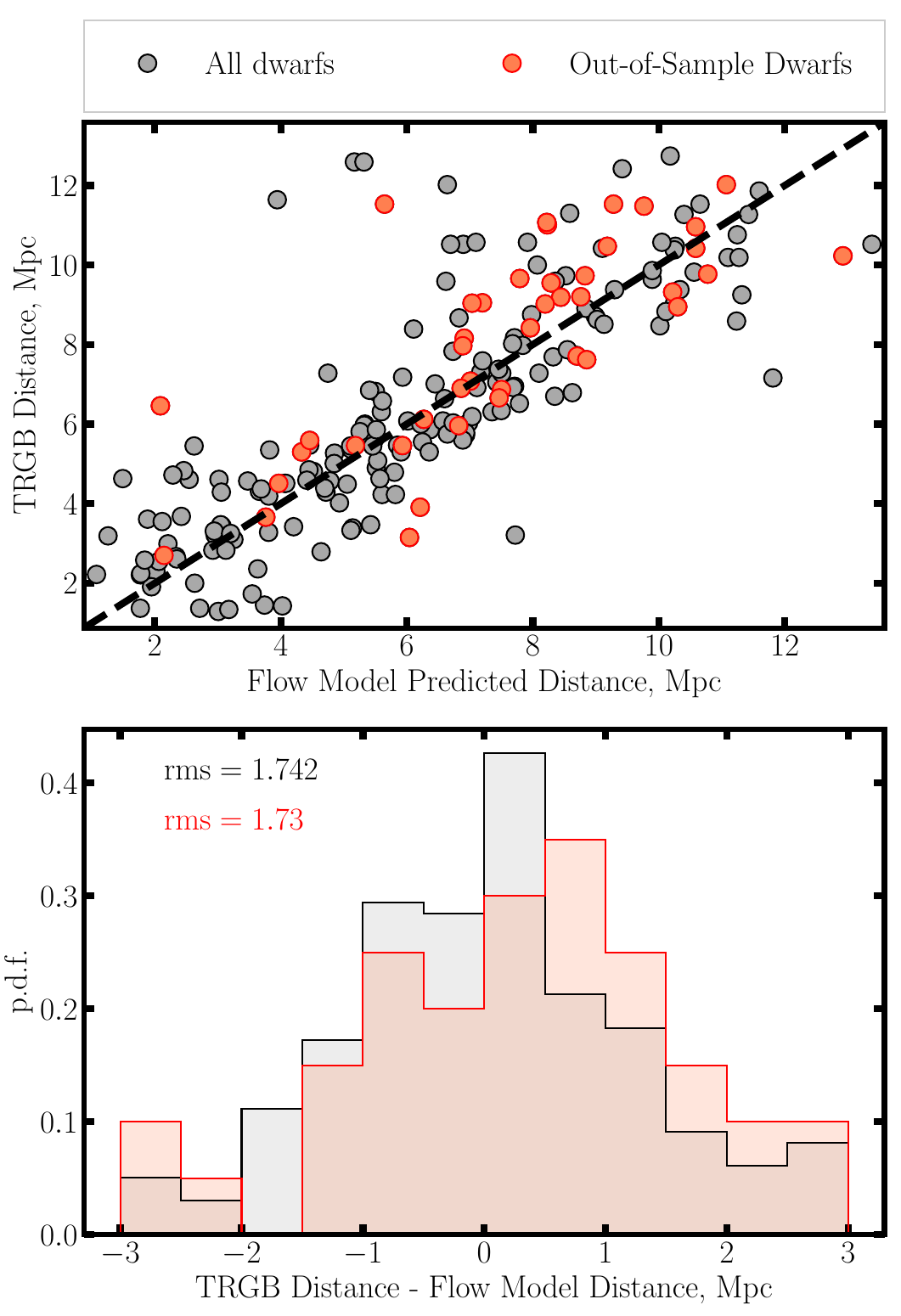}
\caption{A demonstration of the redshift-distance calculator of \citet{kourkchi2020} for a sample of LV galaxies with known TRGB distances from \citet{karachentsev2013}. Since many in this sample were used to calibrate the model of \citet{kourkchi2020}, there is concern that this performance would not generalize to new galaxies. To test this, the red points show LV galaxies that were not included in the calibration of the \citet{kourkchi2020} model. The predictions of the redshift-distance calculator for these galaxies are of a similar accuracy to the rest of the sample. The bottom panel shows the approximately Gaussian residuals.}
\label{fig:redshift}
\end{figure}

To quantify how well this distance calculator predicts distances based on redshift and, hence, assign a realistic error to the distance, we compare its predictions with TRGB distances for the LVGC sample with TRGB measurements. Due to the issues of the redshift-distance calculator for satellites of nearby hosts, we restrict the LVGC sample to those $>500$ kpc projected from a known $M_\star>10^{10}$ \msun\; group. Additionally, we remove any galaxy within $15^\circ$ of the Virgo Cluster. Figure \ref{fig:redshift} shows the agreement between the TRGB distance and the redshift-inferred distance\footnote{Note that we are not restricting the LVGC sample to the ELVES-Field footprint, and that is why there are a couple hundred points as opposed to the roughly two dozen dwarfs with TRGB measurements in the footprint (cf. \S\ref{sec:trgb}).}. The agreement is generally good. For all of the objects that are known to be nearby, the redshift-distance calculator also estimates them to be $D\lesssim 12$ Mpc. Admittedly at an individual galaxy level, some galaxies have redshift-based distances that are factors of multiple off from the TRGB distance \citep[see e.g.][for other examples of this]{mcquinn2021}, but at the population level, this redshift-distance estimator appears to be an unbiased albeit noisy distance metric. This result is similar to the recent finding of \citet{karachentsev2024} who also compared TRGB distances to the redshift-distance calculator of \citet{kourkchi2020} for the LVGC sample.

Since the \citet{kourkchi2020} calculator is based ultimately on Cosmic-Flows3 which will include many of the TRGB distances in this sample, there is concern that this is not a fair test. To explore this, we separately indicate LVGC galaxies that had their TRGB distances measured since the publication of \citet{kourkchi2020}. In particular, we flag galaxies who have TRGB distances currently reported in the LVGC but did not in a version of the catalog dated from May of 2020. These are labeled as `out-of-sample' dwarfs in Figure \ref{fig:redshift}. The result for these galaxies is very similar to that of the entire sample. 

The bottom panel of Figure \ref{fig:redshift} shows the residuals of the redshift-distance calculator compared to the TRGB distances. The residuals appear approximately Gaussian with $\mathrm{r.m.s}\approx1.75$ Mpc. Based on this, we simply assign a 2 Mpc error to the distances returned by the calculator. In the end, 498 ($\sim20\%$) of the candidates have existing redshift information.

\subsection{Surface Brightness Fluctuations}
\label{sec:sbf}
Due to the sparsity of existing TRGB or redshift measurements for the candidates, the majority of the distance constraints come from new SBF measurements. Our SBF process is based on that done in the ELVES Survey \citep{carlsten2020b, carlsten2022}, so we only briefly summarize here and highlight the changes to the procedure.

SBF involves measuring how `semi-resolved' a galaxy is in imaging data. Due to Poisson fluctuations in the number of bright, giant stars in each resolution element of the image, a galaxy can look `patchy' or `mottled' and will appear smoother the further away it is. Measurement of SBF is simply a quantification of this level of patchiness. We start with a smooth model of the galaxy's brightness profile in the HSC imaging. In cases where we have both $r$ and $i$ band coverage in HSC, a judgment is made based on the depth and PSF-size in the two bands. Due to SBF being intrinsically stronger in redder bands, we almost always prefer $i$ band over $r$ unless the $r$ band is significantly deeper. We manually fit a S\'{e}rsic profile to the HSC data to use as this smooth model. We do not simply use the same S\'{e}rsic fit from the detection step, since the HSC data is generally significantly deeper and we do not use any $i$ band data from LSDR10. Then, this model is subtracted from the original image and the result is normalized by the square root of the model. Via Fourier transform, we calculate the power spectrum of this residual. The SBF component of the fluctuations (as opposed to white noise in the image) is the component of the power spectrum on the scale of the PSF. The PSF is modeled with \texttt{PSFEx} \citep{psfex}.

\subsubsection{Mask and Annulus Choice}
As discussed in \citet{sbf_calib}, the choice of mask and annulus used to calculate the power spectrum is quite important and is often a source of ambiguity in the SBF measurement. It is critical to mask out foreground stars, background galaxies, and/or clusters and star-forming clumps from within the galaxy itself. Left unmasked, these contaminants could utterly dominate and bias the SBF measurement \citep[e.g.][]{mei2005, blakeslee2009}. However, for many of the closer galaxies, the SBF signal itself can be quite pronounced, and one runs the risk of masking peaks in the SBF if using too aggressive of a mask. Similarly, the annulus must be large enough to contain a statistical number of resolution elements but not so large that it lets in  noise from the outskirts of the galaxy and ruins the signal-to-noise\footnote{We note that machine learning-based methods using convolutional neural nets might obviate the need for specific mask and/or annulus choices \citep[e.g. \textsc{Silkscreen},][]{miller2025}, but we opt to use more `classic' SBF methodologies here.}.

In this work, we take a slightly different approach from the ELVES Survey and try to systematize the mask and annulus choice relative to that work. For the annulus, we find that defining the annulus in terms of physical surface brightness thresholds works for a wide variety of dwarfs (i.e bright vs. faint and blue vs. red). We simply define the annulus to be the region where the smooth galaxy model is between 23 and 25 mag arcsec$^{-2}$. Since very LSB targets might have central SB that are even lower than 25 mag arcsec$^{-2}$, we also institute a minimum annulus area in pixels that ranges from 7000 pixels to 2000 pixels based on the integrated magnitude of the galaxy. In the case where an annulus falls below this minimum size, the outer edge is widened until the minimum size is reached. This definition of the annulus works particularly well for brighter (and often irregular) dwarf candidates. This range in SB generally encompasses the outskirts of these dwarfs which are often far rounder and more regular than the irregular central regions \citep[see e.g.][for discussion of the science behind this]{kadofong2020, carlsten2021a}.

For the mask, we attempted to devise a system that allowed for very aggressive masking when SBF was not present but wouldn't overmask in the cases where significant SBF was exhibited. To do this, we simply use a masking threshold that is tied to the amount of variance in the annulus. In particular, we mask pixels of the annulus that are $>3\times$ the median absolute deviation (MAD) of pixels within the annulus. The rationale is that if SBF is present, the MAD will be higher to reflect this, and peaks in the SBF will not be masked. Alternatively, if SBF is not present but only a few contaminating point sources are present, the MAD should be lower and the mask will be more aggressive. 

We find that the masks and annuli chosen in these ways work for the majority of candidates. There are cases where the masking prescription overmasks the SBF peaks in very nearby (i.e. almost fully resolved) dwarfs. We manually fix these cases when they appear by raising the masking threshold. Occasionally, the masking threshold is manually lowered when there are clear (non-SBF) contaminating sources that are not getting masked. The annuli chosen by the above prescription almost never need to be changed. In total, we manually adjust the mask and/or annulus in $12\%$ of cases. In most of these cases, the adjustment is clearly needed and the mask and/or annulus choice is unambiguous. However, in a handful of cases ($\sim5$\% of adjustments) the mask choice is ambiguous to the point of materially changing the inferred SBF distance whereby a candidate could be considered a LV dwarf or not, depending on the mask threshold. In these, we use our best judgment in selecting a mask threshold but flag these candidates as having potentially biased SBF distance results. Due to the small number of these problematic candidates, our main science conclusions do not change if the masking thresholds we used turn out to be inappropriate.

\subsubsection{SBF Measurement}
With the mask and annulus in hand, we measure the SBF variance for each candidate from the power spectrum. We fit the power spectrum over the range in wavenumber $k \in [0.05,0.4]$ pixel$^{-1}$ with a combination of a contribution from white noise and a contribution from fluctuations on the scale of the PSF. The measured SBF is the latter component. Unlike some dwarf SBF procedures that vary the wavenumbers included in the fit \citep[e.g.][]{cohen2018}, we simply fix the range used. % but do try different ranges in $k$ and flag the galaxy if the SBF result is very sensitive to this (see below). 
Due to the difficulty in propagating pixel uncertainties through the Fourier transform and into the power spectrum, we estimate the statistical uncertainty on the SBF measurement via Monte Carlo. After measuring an SBF variance level from the image, we do 100 image simulations injecting SBF at the measured level onto the smooth galaxy model within an empty noise field. The noise field comes from the inverse variance maps produced by the HSC pipeline. We then quantify how well we can recover this SBF variance. The spread is used as an estimate of the statistical uncertainty in the SBF measurement. The recovered SBF variances show an approximately Gaussian distribution. If the recovered SBF signal from the simulations shows a systematic bias, the measured SBF variance is corrected for this. This is often found to exist at the $\sim10\%$ level. Following \citet{sbf_calib}, we also measure the SBF signal in a grid of nearby background fields in the HSC imaging. We subtract the median measured SBF power from these fields and incorporate their spread into the uncertainty in the SBF measurement. These background fields account for the average amount of fluctuation power coming from unresolved, faint contaminating sources below the masking threshold.

\subsubsection{Calibration and Final Distance Measurement}
\label{sec:sbf_calibration}
The measured SBF fluctuation power is converted into a distance using the color-based calibration from \citet{sbf_calib} for $i$ band HSC data and that from \citet{li2025} for $r$ band HSC data. Both of these calibrations model the absolute SBF power as a function of galaxy color for a sample of reference dwarfs with known TRGB distances. This accounts for the impact of stellar population on the measured SBF fluctuation power as bluer galaxies (e.g. younger and/or less metal-rich) will exhibit stronger SBF at fixed distance. In recent years, a couple of other empirical SBF calibrations specifically meant for the blue colors of dwarf galaxies have been published \citep{kim2021, cantiello2024}. These calibrations generally show good agreement with each other and with the expectations from stellar population synthesis models \citep{greco2020}. The galaxy color that we use in the calibration is always a $g-r$ color measured from the LSDR10 images using the same annulus as used for the SBF measurement. This color can be different from the integrated color coming from the S\'{e}rsic fit as many dwarfs have outskirts that are redder than their central regions.  

The \citet{sbf_calib} and \citet{li2025} calibrations are specifically for the CFHT/MegaCam filter system while we are measuring SBF in HSC filters and color in DECam filters. To quantify any additional error from filter differences, we performed experiments calculating theoretical SBF magnitudes and colors using the MIST isochrones and synthetic photometry \citep{mist_models} for MegaCam, HSC, and DECam. We find the SBF magnitudes are always quite similar between MegaCam and HSC filters (both $r$ and $i$) with differences $\lesssim 0.06$ mag ($\sim 3\%$ effect in distance). Thus, we do not perform any adjustments to the measured SBF magnitudes. We find a small but meaningful difference between MegaCam and DECam $g-r$ colors, however. We convert the measured DECam colors into the MegaCam filter system with $(g-r)_{\mathrm{MegaCam}} = 0.908\times(g-r)_{\mathrm{DECam}} - 0.002$. Due to its inclusion in the calibration, the uncertainty in the galaxy color is an important component of the final SBF distance uncertainty. We estimate an uncertainty in color using artificial dwarf injection simulations similar to those discussed in  \S\ref{sec:completeness}. Using these simulations, we quantify how well we recover galaxy color as a function of dwarf $g$ magnitude and use this to estimate uncertainties for actual dwarf candidates. 

The error in the measured SBF fluctuation power is roughly Gaussian distributed, but this will not necessarily correspond to a symmetric, Gaussian distributed error in physical distance. We estimate the uncertainty in the final distance using Monte Carlo, incorporating the uncertainty in the fluctuation power and color, and record the 2.5th, 16th, 84th, and 97.5th percentiles in the distance distribution. Below, when we refer to the $2 \sigma$ SBF distance lower bound, we use the 2.5th percentile in the distance p.d.f.

\subsubsection{SBF Quality Check}
A final component of the SBF measurement process is a flag we define to help identify candidates whose SBF distances are possibly spurious. In particular, we want to identify candidates whose measured SBF is likely coming from some source other than genuine SBF, such as irregular morphology or residuals to the S\'{e}rsic fit. For these candidates, the SBF distance will be an underestimate of the true distance. This SBF `trustworthiness' flag has three main components. First, the residuals of the power spectrum fit at low wavenumber (i.e. at large scales) are quantified. If the average residual between $0.02 < k < 0.05$ pixel$^{-1}$ is above a certain threshold, it indicates that a significant amount of the power is possibly coming from large-scale irregularities, likely residuals from the S\'{e}rsic fit, and not genuine SBF. We set this threshold after visually inspecting a large number of candidates and their SBF power spectrum fits. Second, we split the chosen measurement annulus in half and perform the SBF measurement independently for the inner and outer sub-annuli. If they differ by more than $1.5\sigma$, it indicates that the measurement is particularly sensitive to the chosen annulus and should be viewed with caution. Finally, we consider the SBF measurement of any candidate with ellipticity $>0.75$ to be suspect since galaxies that are that flattened will very likely have significant residuals to the S\'{e}rsic fit. A candidate that satisfies any of these three criteria is flagged to have an untrustworthy SBF measurement, and for these galaxies, we can not use SBF to confirm them as LV members. However, for these flagged candidates, we still use the SBF distance to set a valid distance lower bound. The measurement problems that this flag is designed to indicate would all bias the SBF distance to be too low by adding spurious fluctuation power to the image. Thus, even with these issues, the SBF measurement can set a conservative lower bound on the distance. Note that this `trustworthiness' flag is separate from the flag described above, which indicates ambiguity in the masking threshold choice. This `trustworthiness' flag is set to \texttt{False} for many more candidates.

Since there are some significant differences between the SBF methodology used here and that used in the ELVES Survey, we tested it using both SBF image simulations and on real galaxies in the HSC footprint that have known distance. We go into detail on these tests in Appendix \ref{app:sbf_tests} demonstrating that the current, more systematic, SBF methodology agrees well with known TRGB and redshift distances as well as SBF distances published in ELVES.

\subsection{Dwarf Confirmation}
\label{sec:dist_results}
Using the distance information available, we confirm or reject candidates as genuine LV dwarfs in a similar way as in the ELVES Survey. A candidate whose $2\sigma$ distance lower bound is within 10 Mpc is considered a `confirmed' LV member while those with lower bounds beyond 10 Mpc are deemed `rejected' candidates. For candidates with distance information from multiple sources (e.g. redshift and SBF), we consider TRGB distance information with the highest priority, then SBF, and then redshift. TRGB and SBF distances are always consistent within their measurement uncertainties. On the other hand, redshift distances are occasionally discrepant from TRGB distances and/or SBF distances. In many of these cases, the dwarf is in a virialized region around a massive host (e.g. an ELVES host or massive background group), but in many other cases, the redshift-based distance appears to simply be spurious (see the scatter in Figure \ref{fig:redshift}).% TRGB and redshift-based distances disagree in their confirm/reject classification in only \todo{xx}\% cases while SBF and redshift disagree in only \todo{xx}\% cases, indicating overall fair agreement between the different distance metrics. 

When using the SBF distances, for a candidate to be `confirmed' as a LV dwarf, two further criteria must be met. First, we require that the SBF `trustworthiness' flag described in \S\ref{sec:sbf} be set to \texttt{True}. Second, following ELVES, the SBF measurement must be of high signal-to-noise (S/N) with S/N $>5$. We use this threshold of S/N $>5$ as a safeguard against false positive dwarf confirmation. In our experience with the ELVES Survey, this S/N level is roughly where SBF becomes clearly distinguishable from other sources of fluctuation power, such as scattered star-forming clumps in the galaxy or background galaxies. As part of this safeguard, we  closely visually inspect the HSC cutouts to confirm that the measured fluctuation power appears to come from bona fide SBF. For a small number ($28$) of candidates with high S/N$>5$ SBF measurements, it appears that the fluctuation power is due instead to non-SBF sources, such as clumps in the galaxy or subtle irregular morphology, and we manually fail this check for these galaxies. 

For candidates where either criterion is not met, we only consider the $2\sigma$ distance lower bound. If that is beyond 10 Mpc, we still reject the candidate. Failing these criteria means that much or all of the measured fluctuation power for a given candidate is coming from some source other than genuine SBF. Thus the distance lower bound will be conservative and it is valid to reject a candidate using it. However, if the lower bound is within 10 Mpc, we remain agnostic about the dwarf and deem it an `unconfirmed' candidate. We describe our treatment of these in \S\ref{sec:maybes} below.

For confirmed dwarfs with multiple distance methods, to define a single distance and associated uncertainty, we again use any TRGB distance with the highest priority, then SBF, then the redshift-based distance. As mentioned in \S\ref{sec:sbf_calibration}, the SBF distance uncertainty is somewhat asymmetric with the $+1\sigma$ uncertainty generally slightly larger than the $-1\sigma$. However, this difference is $\lesssim 20\%$ and so, for simplicity, we average the SBF $+1\sigma$ and $-1\sigma$ estimates and report only a single, symmetric distance errorbar.

In many of the analyses below, we define a `confirmed weight' that incorporates the likelihood that the dwarf is truly within 10 Mpc given its measured distance and uncertainty. Thus, a `confirmed' LV dwarf with a measured distance of $10\pm1.5$ Mpc will have a confirmed weight of 0.5. This will be the weight that the dwarf contributes to, for example, the field luminosity function. How unconfirmed candidates get treated and included in, for example, the luminosity function is discussed below.

Appendix \ref{app:photometry} includes the final list of dwarfs and candidates found in ELVES-Field. In total, there are 95 confirmed dwarfs and 600 remaining unconfirmed/candidate dwarfs. In addition to confirmed and candidate dwarfs, we also list the locations of detections that were rejected from some distance measurement, of which there were 1639.

\begin{figure*}
\includegraphics[width=\textwidth]{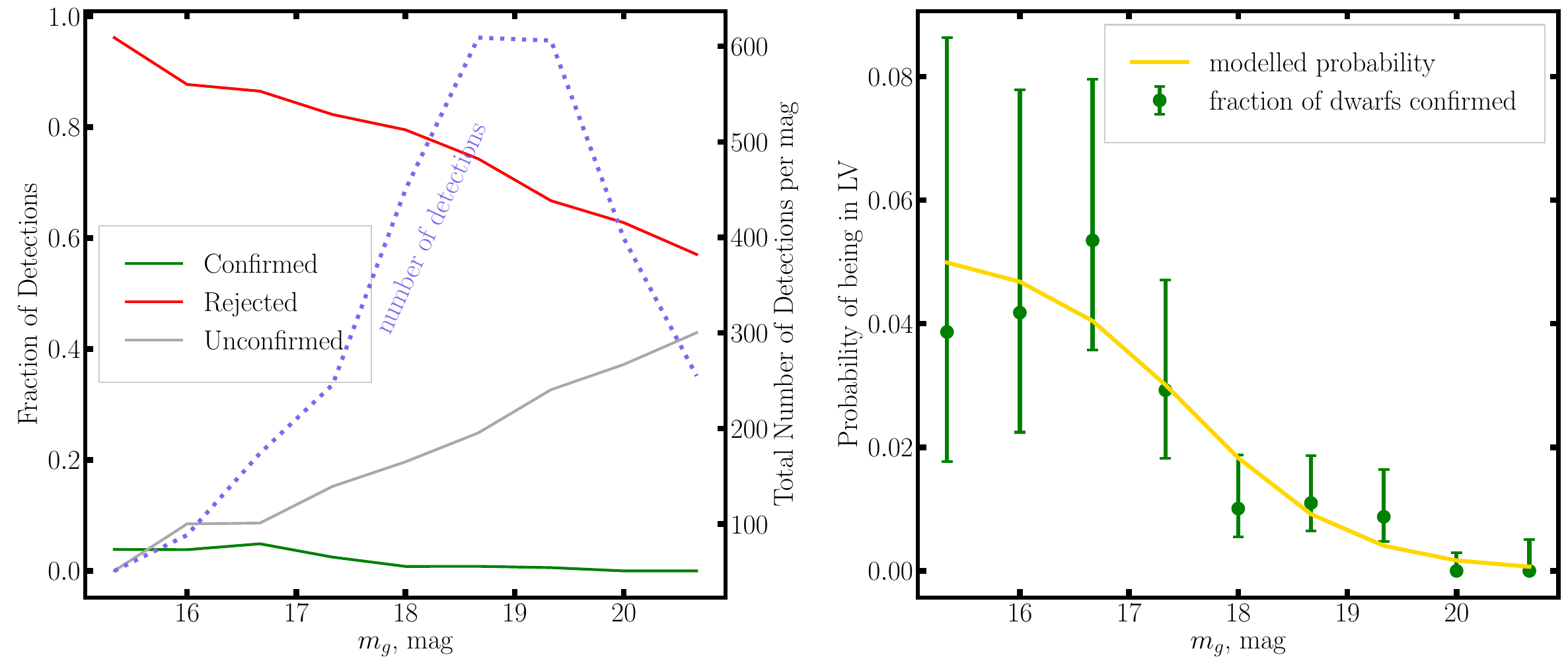}
\caption{The left panel shows the number of candidate, isolated dwarf detections as a function of $g$ magnitude as well as the fraction of these candidates that get confirmed or rejected as genuine LV dwarfs. `Unconfirmed' candidates did not have a successful distance measurement, and their status as true LV dwarfs remain unknown. The right panel shows the fraction of dwarfs that got confirmed as LV dwarfs out of the total that did have a successful distance measurement. We use this fraction as an estimate for the probability that a given unconfirmed dwarf without distance information is a real LV dwarf based on its $g$ magnitude. The green points are the observed fraction of confirmed dwarfs in each magnitude bin, and the gold line is a sigmoid function that is fit to the points.   }
\label{fig:psat}
\end{figure*}

\subsection{Unconfirmed Dwarf Candidates}
\label{sec:maybes}
After applying the available distance information, a candidate dwarf will be categorized into one of three groups: 1) confirmed LV dwarf, 2) rejected background contaminant, or 3) an `unconfirmed' candidate. The third group contains detections where the SBF measurement was inconclusive (e.g. due to too low signal-to-noise). These objects may or may not be genuine LV dwarfs and will need deeper data or, most likely, space-based observations to measure their distances. However, using the properties of the confirmed and rejected candidates, we can estimate the probability that each of these unconfirmed candidates is a true LV dwarf. This is a similar process as used in the ELVES and SAGA Surveys for candidate satellites that did not have successful distance measurements.  

Figure \ref{fig:psat} shows the number of total candidate detections, as well as the fraction of candidates in each of the three groups (confirmed, rejected, and unconfirmed) as a function of $g$ magnitude. The confirmed dwarfs contribute to the histogram according to their `confirmed weight' as described at the end of \S\ref{sec:dist_results}. The fraction of unconfirmed candidates increases at fainter magnitudes due to an increased likelihood of low S/N in the SBF measurement. The right panel shows the fraction of dwarfs that get confirmed out of all those that have successful distance measurements (i.e. the confirmed and rejected candidates). The green errorbars show the actual measured fraction with confidence intervals estimated by treating the fraction as a binomial parameter. We use this fraction as an estimate of the probability that an unconfirmed candidate is an actual LV dwarf. In particular, we fit a sigmoid function to the measured confirmed fraction and use that to estimate the probability that an unconfirmed candidate is in the LV  based on its $g$ magnitude. This sigmoid fit is shown as the gold line in Figure \ref{fig:psat}. Due to the vast majority of detections being background contaminants, this probability is always $\lesssim5\%$ and drops below $1\%$ for the faintest candidates. 

In developing this model, we explored the use of other observables, including size, surface brightness, and color, in predicting the probability that an unconfirmed candidate is a real LV dwarf but found that these other variables added only slight, if any, distinguishing power over using $g$ magnitude alone. This is likely because these other variables are quite correlated with $g$ magnitude. Additionally, with the sample size of only several dozen confirmed LV dwarfs, we do not have the statistics to develop a more complicated model.

The use of the confirmed fraction as an estimate for the probability that an unconfirmed candidate is a real LV dwarf is clearly quite approximate. It assumes that the level of data quality required to confirm an LV dwarf at a given magnitude is the same as that required to reject a background contaminant at that magnitude. This is likely the case for redshift-based confirmation and rejection. We make use of redshifts that come primarily from major wide-field redshift surveys, such as SDSS. These surveys will get redshifts for almost all sources brighter than a certain threshold in apparent magnitude, and thus will be as capable to reject a candidate as to confirm it if the candidate is brighter than that threshold. This assumption is more dubious when using SBF. For instance, we can consider an alternate scenario where we require a S/N $>100$ measurement of the SBF to `confirm' a dwarf (as opposed to the S/N $>5$ threshold actually used, see \S\ref{sec:sbf}). If this was the case, rejecting a background contaminant would be possible with much worse data than that required to confirm the dwarf. Thus the rejected contaminants would be over-represented in the set of candidates with conclusive distance results, and Figure \ref{fig:psat} would underestimate the probability that an unconfirmed candidate is a real LV dwarf. We address this with image simulations in the ELVES Survey \citep{carlsten2022} and find that, for a given quality in imaging data, we are indeed able to reject candidates with SBF down to slightly fainter magnitudes than what we can confirm candidates down to. However, the difference is $\lesssim 1$ mag. Since this is small compared to the range in the magnitudes of the unconfirmed candidates, we expect this probability estimate to be reasonable.

% the below argument is based on this "math" --- D_lb is the measured distance lower bound for a maybe
% P(real | m_g, D_lb) ~ N_real( m_g, D_lb) / N_tot(m_g)  ---- the probability of a maybe being real at a given m_g and D_lb is approximately the fraction of confirmed things at that m_g and satisfying that D_lb out of the total num of candidates at that m_g. 
% but we don't have the number of confirmed things to split them up by m_g *and* D_lb. so we make this approximation: 
% N_real(m_g, D_lb) ~ N_real(m_g) * fraction of space volume( > D_lb)  ---- basically just assuming things are evenly distributed in 3D space.  

The probabilities that we estimate from Figure \ref{fig:psat} do not account for any (albeit weak) distance constraints from the SBF measurement that might exist. For most of the unconfirmed candidates, we have some distance lower bound set from the SBF measurement. It is simply not constraining enough or high enough S/N to fully reject or confirm the candidate. Still this distance information should inform the estimated probability of a candidate being in the LV. Intuitively, an unconfirmed candidate with a distance lower bound of $>9.99$ Mpc at a given $g$ magnitude should have a lower estimated probability of being a real LV dwarf than a candidate at the same magnitude that has a much weaker distance lower bound (e.g. $>1$ Mpc). By assuming dwarfs are uniformly distributed in 3D space, we can incorporate this distance information by multiplying the probability estimated from a candidate's $g$ magnitude by a modifier that represents the fraction of the volume of space in the LV that is beyond that candidate's distance lower bound. For instance, for a candidate with distance lower bound $>9$ Mpc, this modifier would be $9^3 / 10^3 \approx 0.73$.

Finally, we note that for this probability estimation, we have only used `isolated' dwarf candidates. We defer a full discussion of how we define `isolated' until \S\ref{sec:isolation}. In brief, we flag a candidate as being `isolated' if it is outside of a projected $2\times R_{\mathrm{vir}}$ radius from any group with total stellar mass $M_\star > 10^9$ \msun~ within $D<16$ Mpc from the \citet{kourkchi2017} group catalog. We find that the estimated probabilities of the unconfirmed candidates being real LV dwarfs do not change significantly if we don't use this isolation criterion and instead use the entire sample of candidates. We have redone the science analyses below after either doubling or halving these estimated probabilities and find that the results are not particularly sensitive.

\subsection{Summary of Distance Analysis}
\label{sec:distance_summary}
In the end, with the available distance information, we categorize each detection into one of three groupings: 1) confirmed LV dwarf (95 in total), 2) rejected background contaminant (1639 in total), or 3) unconfirmed candidate (i.e. the dwarf might be a true LV dwarf or it might not be, 600 in total). We describe each below:

\begin{itemize}

\item Confirmed LV dwarfs are those candidates that have a robust distance measurement with a $2\sigma$ distance lower bound that is within 10 Mpc. For an SBF distance measurement to count as `robust', and hence be used to confirm a dwarf, it must be high S/N (S/N$>5$) and pass both a visual check and an automated `trustworthiness' flag that checks for various possible pitfalls in the SBF process (see \S\ref{sec:dist_results} for details). Due to our use of distance lower bounds in determining if a dwarf is `confirmed', it is likely that several confirmed LV dwarfs are actually slightly beyond 10 Mpc. To account for this, confirmed dwarfs get a `confirmed weight' which is the chance they actually are within 10 Mpc based on their measured distance and uncertainty, assuming Gaussian uncertainties. This weight is used in many of the analyses below and determines, for example, how much each confirmed dwarf contributes to the average luminosity function.

\item  Rejected background contaminants are those whose $2\sigma$ distance lower bounds are beyond 10 Mpc.

\item Unconfirmed candidates are those for which either there was no SBF measurement performed or the measurement was too low S/N for a conclusive confirmation or rejection. Based on the fraction of dwarfs with conclusive distance measurements that turned out to be real LV dwarfs, we estimate a probability, $P_{LV}$, for each of these candidates indicating the probability that they are true LV dwarfs. We will incorporate these unconfirmed candidates in many analyses below using this estimated probability.

\end{itemize}

\section{Dwarf Catalog}
\label{sec:dwarf_properties}

%\textcolor{red}{MOVE THIS UP BEFORE DISTANCE MAYBE?} -- % i think this being after distance makes sense

\subsection{Dwarf Photometry}
\label{sec:dwarf_photometry}

The photometry for the confirmed and candidate LV dwarfs comes from single S\'{e}rsic fits. As mentioned in \S\ref{sec:algorithm}, the majority of the fits used are those produced systematically in the detection pipeline. However, around $\sim25\%$ of the fits are tweaked manually (e.g. the mask or cutout size is changed) to produce a more visually acceptable fit. Following the ELVES Survey, we estimate uncertainties on the photometric quantities using injection simulations similar to those in \S\ref{sec:completeness}. From these simulations, we model the typical recovery error of different photometric quantities as a function of apparent $g$ band magnitude and use this to assign uncertainties to the detected dwarfs. 

We assign stellar masses to the confirmed LV dwarfs using a relation from \citet{delosreyes2024} based on a dwarf's $g-r$ color and absolute $M_g$ magnitude. The stellar mass uncertainty includes the uncertainties in apparent $g$ magnitude, color, and distance. Due to the requirement of distance, we only assign stellar masses to fully confirmed LV dwarfs. For the remaining unconfirmed candidate dwarfs, we include them in the various analyses below stochastically using their estimated probabilities of being true LV dwarfs. This process will be described further below.

Appendix \ref{app:photometry} includes tables of the photometry for all the confirmed and candidate dwarf detections.

\subsection{Dwarf UV Photometry}
\label{sec:dwarf_uv_photometry}
Archival GALEX \citep{galex} imaging exists for many of the confirmed LV dwarfs and offers an important window into the star-forming properties of the dwarfs. In this, we follow almost exactly the process used in the ELVES Survey which, in turn, was based on the methods of \citet{greco2018_two} and \citet{karunakaran2021}. We only do this process for the confirmed LV dwarfs and not for any of the remaining unconfirmed dwarfs. For each dwarf, we search the MAST archive for GALEX coverage. We find at least some coverage in NUV for a majority of the confirmed dwarfs, 88/95. About a third of these have coverage from the shallow GALEX All-sky Imaging Survey, while the remainder have deeper data, often from the Nearby Galaxy Survey \citep{galex2007}. Due to the shallowness of GALEX FUV data and the faintness of most of the dwarfs in the FUV, we focus only on NUV data here.

We refer the reader to \citet{carlsten2022} for the full details on how we extract the GALEX photometry. In brief, we perform aperture photometry using elliptical apertures of radii twice the effective radii found from the LSDR10 S\'{e}rsic fits. Contaminating point sources are masked with a \texttt{SExtractor} \citep{sextractor} run. The sky contribution to the flux is estimated from the median of 50 apertures placed in the vicinity of the galaxy within the GALEX frame. The dispersion of sky values measured in these apertures is used in calculating the flux uncertainty and setting flux upper limits when the galaxy is not detected in GALEX. We use the zeropoints of \citet{morrissey2007} to convert the measured fluxes to AB magnitudes and correct for Galactic extinction using $R_{\rm NUV}=8.2$ \citep{wyder2007}. For dwarfs that had a S/N$<2$ detection, we report $2\sigma$ upper limits to the NUV flux. 

Appendix \ref{app:photometry} includes a table of the NUV photometry for all the confirmed LV dwarfs in ELVES-Field.

\subsection{Dwarf Environment}
\label{sec:isolation}
Since the goal of the ELVES-Field Survey is to characterize isolated nearby dwarf galaxies, defining `isolated' clearly plays an important role. We base our definition of environment on the group catalog of \citet{kourkchi2017}. This catalog is based on the Cosmic-Flows3 \citep{tully2016} catalog of nearby galaxy distances and thus will be the most accurate group catalog for nearby groups. From this catalog, we select all groups with total stellar mass $>10^9$ \msun~ and distance $<16$ Mpc. The group stellar mass is calculated from the group total $K_s$ band luminosity assuming a mass-to-light ratio of $M_\star/L_{K_s}=0.6$ \citep{mcgaugh2014}. In cases where the group catalog does not list a Cosmic-Flows3 distance for a group, we use the estimated distance from the redshift-distance calculator of \citet{kourkchi2020}. Using the estimated `second turnaround radius' as a proxy for the group virial radius, we flag detected dwarfs if they fall within either a projected $1\times R_{\mathrm{vir}}$, $2\times R_{\mathrm{vir}}$, or $3\times R_{\mathrm{vir}}$ from any \citet{kourkchi2017} group. These different flags represent different levels of isolation with the dwarfs being outside of $3\times R_{\mathrm{vir}}$ of any group being the most isolated. We use the $2\times R_{\mathrm{vir}}$ definition of isolated as the fiducial criterion in much of the analysis below.

Using projected separation to determine environment instead of 3D separation and including groups out to 16 Mpc make the isolation label conservative and more resistant to distance errors (both of the group and of the detected dwarf). In fact, the measured distances to the dwarfs are not used in classifying their environments. In using projected separation instead of 3D distances to massive hosts, we expect to miss many truly isolated dwarfs, classifying them instead as `non-isolated'. To quantify what fraction of truly isolated dwarfs are missed in this way, we use the Illustris TNG cosmological simulations to select subhalos within 10 Mpc of MW analogs (see \S\ref{sec_abund_sims} below for more details on the Illustris simulations) and classify them as isolated or not depending on proximity to $M_\star>10^9$ \msun~ host galaxies using both projected separation and 3D distance. For our fiducial $2\times R_{\mathrm{vir}}$ definition of isolation, we find that $\sim30\%$ of truly isolated subhalos fall nearby in projection to a massive host and would be considered `non-isolated' when using projected separation. This has the effect of reducing the sample size of isolated dwarf but ensures the purity of the isolated sample as we expect essentially no false positives in the sample.

The distance upper limit of 16 Mpc is chosen so that most of the groups near to the Virgo Cluster would be included as potential hosts.  The lower limit in stellar mass is about that of the Large Magellanic Cloud (LMC), and thus we are considering satellites of LMC-mass dwarfs to not be isolated as there is some evidence that hosts of this mass might be capable of environmentally quenching their satellites \citep{garling2020, jahn2022}. Importantly, dwarfs flagged as `isolated' might still be satellites of lower mass hosts (i.e. analogous to the Small Magellanic Cloud). Indeed there is some overlap between the sample of `isolated' field dwarfs and the satellites found in the ELVES-Dwarf Survey \citep{li2025}. Finally, we exclude the MW and M31 from the \citet{kourkchi2017} group list because they would be problematic with the use of projected separations. We flag any dwarf that is $D<2$ Mpc as non-isolated.

In addition to the $R_{\mathrm{vir}}$-based isolation flags, we also define an isolation flag based on physical projected distance from a nearby massive host. Following \citet{geha2012}, we flag detected dwarfs that are within 1.5 projected Mpc from any massive galaxy with $M_\star > 2.5\times10^{10}$ \msun. To get a list of potential massive hosts galaxies, we use the galaxy catalog of \citet{kourkchi2017} and again restrict it to galaxies with $D<16$ Mpc. When the galaxy catalog does not list a Cosmic-Flows3 distance for a given galaxy, we fill in the distance from the redshift-distance calculator of \citet{kourkchi2020}.

In the tables of confirmed and candidate dwarfs, we provide the inferred environment flags but do not make any cut on isolation so that users can apply their own isolation criteria as needed for their analyses.

\section{Luminosity and Stellar Mass Functions}
\label{sec:abundance}
In this section, we present one of the main observational results of the ELVES-Field Survey: the abundance of isolated dwarfs in the LV. We first show the luminosity function (LF) and stellar mass function (SMF) of the dwarfs and then do a brief comparison of the observed SMF with that predicted from the Illustris-TNG \citep{tng1} hydrodynamic simulation.

\subsection{Observed Abundance}
\label{sec_abund_obs}

\begin{figure}
\includegraphics[width=0.48\textwidth]{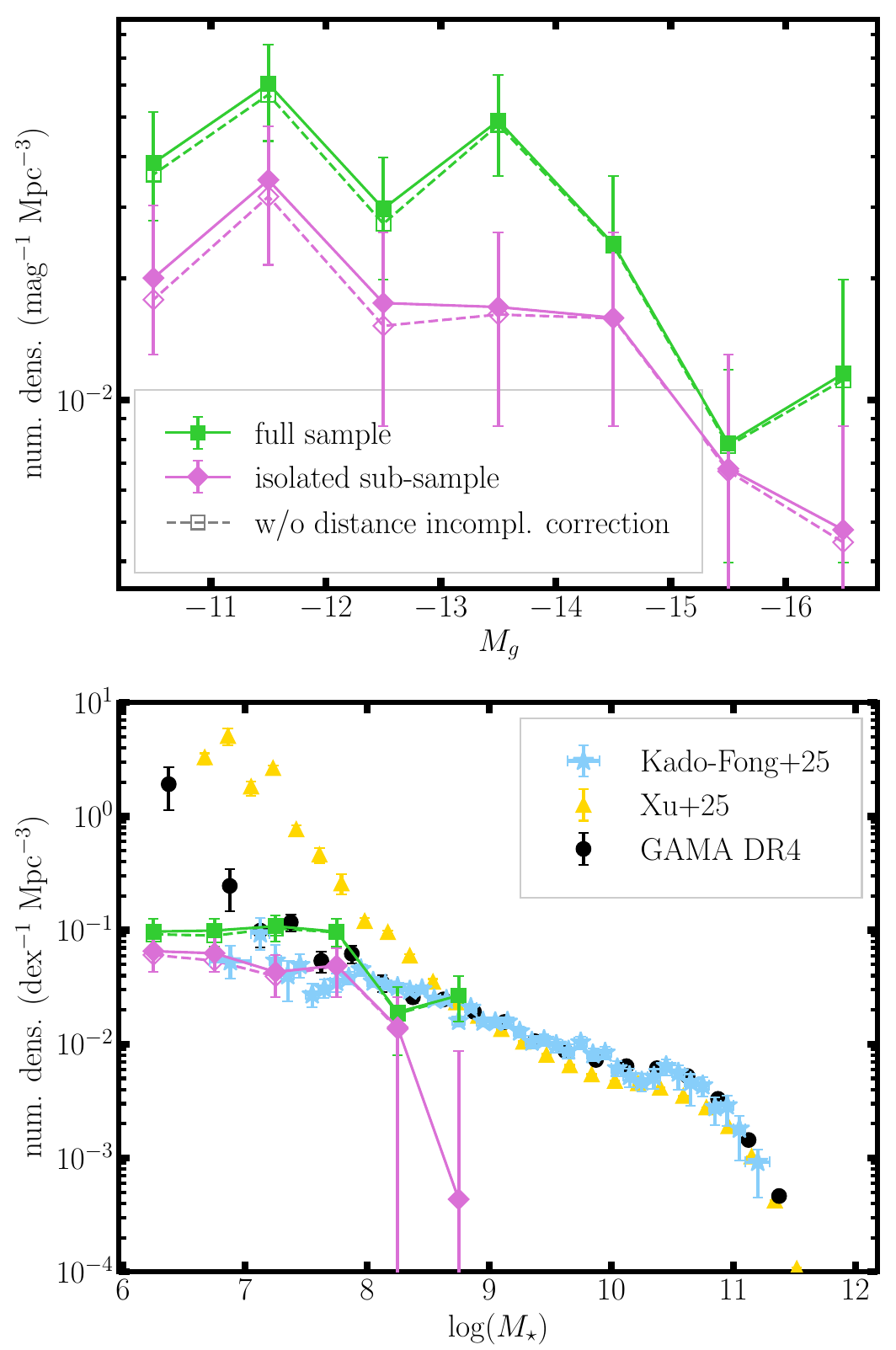}
\caption{The top panel shows the luminosity function of LV dwarfs found in ELVES-Field. We show the LF for both the entire sample of confirmed dwarfs and the sub-sample of isolated dwarfs. Error bars are simply the Poisson uncertainties from the number of dwarfs in each luminosity bin. We include a correction for candidate dwarfs without distance measurements using the estimated probabilities of them being true LV dwarfs from \S\ref{sec:maybes}. The bottom panel shows the analogous stellar mass functions. For comparison, we show the SMF from several recent studies. From our completeness tests, the ELVES-Field sample is at least $80\%$ complete down to $M_\star \sim 10^{6.5}$ \msun~, and thus we should be $>80\%$ complete at the magnitudes and stellar masses shown here. For massive dwarfs, there is good agreement in the abundance across the surveys, although at the lowest masses, GAMA and \citet{xu2025} find a noticeably higher abundance of dwarfs.  }
\label{fig:smf}
\end{figure}

The top panel of Figure \ref{fig:smf} shows the LF of LV dwarfs found in ELVES-Field. We show both the LFs for the entire sample of detected dwarfs in the footprint and the sub-sample of isolated dwarfs. For this, we use the fiducial isolation definition and exclude any dwarfs within $2\times R_{\mathrm{vir}}$ of a massive group from \citet{kourkchi2017}, as described in \S\ref{sec:isolation}. To calculate these LFs, we simply count up the confirmed LV dwarfs in each magnitude bin and normalize by the volume probed in the survey. For this volume, we calculate the fraction of the sky covered in the footprint and multiply by the volume of a sphere out to 10 Mpc. For the complete sample of dwarfs, the area searched in LSDR10 bricks is 2988.9 sq. deg. (accounting for the area excluded around the Virgo Cluster and NGC 5846 group, cf. Figure \ref{fig:survey_area}). For the isolated sub-sample, we recalculate the footprint area after excluding the areas on the sky nearby to massive groups from \citet{kourkchi2017}. This area comes out to be 2746.5 sq. deg. Both of these footprint areas are then further adjusted for the fact that the HSC footprint coverage is smaller than the LSDR10 area. As mentioned in \S\ref{sec:survey_design}, this is due to chip gaps in the HSC data and the fact that we include all LSDR10 bricks that have any overlap with the HSC data. The fraction of candidates detected in the complete LSDR10 footprint that have HSC coverage is $83\%$. Thus we scale down the areas listed above by this factor. Each confirmed dwarf counts towards the LF according to its `confirmed weight', as mentioned in \S\ref{sec:dist_results}. This weight accounts for the uncertainty in the distance measurement used to confirm the dwarf as an LV member.

We also account for the contribution of the remaining unconfirmed candidate dwarfs without conclusive distance measurements in the LFs shown. Using Monte Carlo over many iterations, we calculate the average expected contribution of these unconfirmed candidates to the overall LF using their estimated probabilities of being real LV dwarfs from \S\ref{sec:maybes}. In each iteration, a given candidate is included or not on the basis of this estimated probability, and then a distance is randomly assigned according to any distance lower bound set from the SBF. In Figure \ref{fig:smf}, we display the LFs without this correction as well to show its magnitude.  The contribution to the LF from these unconfirmed candidates is relatively minor ($\sim10\%$) due to the small probability that any given candidate is a real LV dwarf (cf. Figure \ref{fig:psat}).

The errorbars are simply calculated from Poisson statistics on the number of dwarfs in each magnitude bin. We do not account for any effects of cosmic variance nor Eddington bias. Eddington bias would likely cause the LFs shown to be overestimates of the true LFs since dwarfs beyond 10 Mpc will be more numerous than those within 10 Mpc and, thus, more false positive dwarfs will scatter into the LV due to distance measurement errors than true LV dwarfs will scatter out of the 10 Mpc volume we consider. Since the ELVES-Field sample is $>80\%$ complete to $M_\star \sim 10^{6.5}$ \msun~ (cf. Figure \ref{fig:completeness2}) which corresponds to $M_g \sim -11$ mag, we do not need to account for Malmquist bias or other incompleteness effects at the luminosities shown on this plot.

The bottom panel of Figure \ref{fig:smf} shows the corresponding SMFs for the entire sample and sub-sample of isolated dwarfs. These SMFs are calculated in a very similar way to the LFs, including incorporating a contribution from the unconfirmed candidate dwarfs. 

Using different isolation criteria has the effect of essentially scaling up or down the LF and SMF depending on whether the isolation criterion is more restrictive or not. Excluding dwarfs within $1\times R_{\mathrm{vir}}$ of a massive ($M_\star > 10^9 $~\msun) group from \citet{kourkchi2017} instead of $2\times R_{\mathrm{vir}}$ raises the SMF by $\sim25\%$. Excluding dwarfs within $3\times R_{\mathrm{vir}}$ from a massive group scales the SMF down by $\sim20\%$. Using the \citet{geha2012} isolation criterion instead results in approximately the same SMF as the fiducial $2\times R_{\mathrm{vir}}$ criterion.

For reference, we include the SMFs from other recent studies in the literature. These include the GAMA redshift survey \citep{driver2022}, the SAGA-Background redshift survey \citep{kadofong2025}, and a photometric analysis of the Legacy Surveys DR9 \citep{xu2025}. All of these literature SMFs are not environmentally restricted and therefore should be compared primarily to the entire ELVES-Field sample and not the isolated sub-sample. These samples, like the entire ELVES-Field sample, will naturally include a mix of field and satellite dwarfs. \citet{kadofong2025} use the same stellar mass prescription from \citet{delosreyes2024} that we use and is thus the most directly comparable. Within the errorbars, there appears to be fair agreement with \citet{kadofong2025} across the mass range probed by ELVES-Field. 

We make no attempt to correct for the significantly different prescription used in calculating stellar mass in GAMA compared to what is used here so a systematic bias of a few tenths of a dex is possible \citep{klein2024, delosreyes2024}. For the \citet{xu2025} results, we use their reported SMF that is calibrated to the GAMA stellar masses. For brighter dwarfs ($M_\star \sim 10^7 - 10^9$ \msun), there is good agreement between GAMA and the ELVES-Field result. At the lowest masses, the GAMA SMF is substantially higher than what we find here, with the \citet{xu2025} SMF being even higher. Figure \ref{fig:completeness2} shows that ELVES-Field should be $>80\%$ complete down to $M_\star=10^{6.5}$ \msun~ where this discrepancy is. Experimenting with the unconfirmed candidates, we find that the ELVES-Field SMF is only able to reach the GAMA SMF at the lowest mass bin if we set the probability of the unconfirmed candidates to be real, $P_{LV}$, all to 1. However, this is clearly unphysical as we find the vast majority of detections to be background, when the data allow for a conclusive determination. 

This discrepancy is perhaps due partly to cosmic variance or small number statistics with ELVES-Field. Regarding cosmic variance, \citet{xu2025} argue that their SMF is higher than that of GAMA \citep{driver2022} due to the MW's purported location in a local under-density on a scale of roughly $D\lesssim 100$ Mpc \citep[e.g.][]{bohringer2020} where most of the low-mass GAMA galaxies are found. Furthermore, on even smaller scales ($D\lesssim20$ Mpc), there is a well-established `Local Void' \citep{tully2008} with the MW located near the edge. It is possible this plays a role in the lower SMF found by ELVES-Field, but we defer a thorough investigation of this to future work.

\subsection{Comparison to Simulations}
\label{sec_abund_sims}
While a thorough comparison with simulation predictions is beyond the scope of this paper, we do an initial comparison here with the predicted field dwarf abundance from the TNG50 project \citep{tng1, tng50_1, tng50_2}. In this section, we describe how we process and forward model the simulations.

We start with the $z=0$ snapshot of subhalos from TNG50 and select MW/M31 analogs from \citet{pillepich2024} as vantage points to mock observe the field dwarf population. We require the MW/M31 analogs to be more than 10 Mpc from any Virgo-like cluster ($M_{\mathrm{halo}} > 10^{14}$ \msun), leaving 177 options. Then, for each of 1000 iterations, we randomly select one of these MW/M31 analogs as the vantage point along with a random orientation. Subhalos are selected within a 16 Mpc sphere of the vantage point, and the LSDR10 footprint (cf. Figure \ref{fig:survey_area}) is applied. For comparisons with the observed isolated dwarf sub-sample, we also apply an area mask excluding the areas around massive ($M_\star > 10^9$ \msun) groups. As with the observations, we excise a projected area corresponding to $2\times R_{\mathrm{vir}}$ around each of these massive groups within 16 Mpc of the vantage point\footnote{With the observations, we use the `second-turnaround radius' from \citet{kourkchi2017} as a proxy for the virial radius of a group. Similarly, for the TNG50 simulations, we estimate the second-turnaround radius in the same way: $R_{2t} = 0.215(M_{\mathrm{vir}}/10^{12})^{1/3}$ \citep{tully2015}.}. With the subhalos selected, we consider two different options for assigning stellar mass to them. First, we simply use the stellar mass that is predicted by the hydrodynamic and star-formation physics of TNG50. Based on the baryonic resolution of $m_b = 8.5\times 10^4$ \msun\ of TNG50, we impose a lower limit of $M_\star = 5\times 10^6$ \msun\ on dwarf stellar mass to ensure each dwarf is resolved \citep{engler2021, engler2023}. Alternatively, we also assign our own stellar masses to the subhalos based on the constant-scatter stellar-to-halo-mass relation (SHMR) from \citet{danieli2023}. This SHMR was fit to reproduce the observed satellite abundance in the ELVES Survey. 

To best mimic the observational treatment of distances, we apply similar criteria to `confirm' dwarfs as being within 10 Mpc of the vantage point. Since the distance uncertainty is quite different between the different distance methods we use (e.g. TRGB vs redshift), we randomly assign a distance uncertainty to each TNG50 dwarf based on the approximate fraction of ELVES-Field candidates confirmed via the different distance measurement methods. In ELVES-Field, $\sim 15\%$ of the confirmations are from TRGB which has a $\sim 5\%$ distance uncertainty, $\sim 20\%$ of confirmations are via redshift with a $\sim2$ Mpc distance uncertainty (Figure \ref{fig:redshift}), and the remaining $\sim 65\%$ of confirmations are via SBF. In the SBF case, to get a realistic distance uncertainty, for each TNG50 dwarf, we randomly take a real, observed LV dwarf confirmed via SBF and apply the estimated distance error (including any asymmetry) from that dwarf to the TNG50 dwarf.

We note that this is an approximation as dwarfs confirmed via redshift in ELVES-Field are generally brighter than those confirmed via SBF. Dwarfs confirmed by TRGB, on the other hand, are fairly evenly distributed in luminosity. However, this will be a minor effect as the distance uncertainties of redshift and SBF are similar. The TNG50 dwarfs then have their distances scattered by these assigned distance uncertainties. We next `confirm' TNG50 dwarfs as being in the analogous `Local Volume' if their $2\sigma$ distance lower bound is within 10 Mpc of the chosen vantage point.

Finally, we apply the detection selection function that we infer for ELVES-Field from the artificial dwarf injection tests (Figure \ref{fig:completeness}). TNG50 dwarfs are assigned a $g$ band magnitude based on the color-mass-to-light ratio relation of \citet{delosreyes2024} assuming a color of $g-r=0.35$ and are assigned an effective radius using the mass-size relation of \citet{carlsten2021a}. They are then randomly detected or not based on the estimated completeness of the ELVES-Field detection pipeline at that size and magnitude.

\begin{figure}
\includegraphics[width=0.48\textwidth]{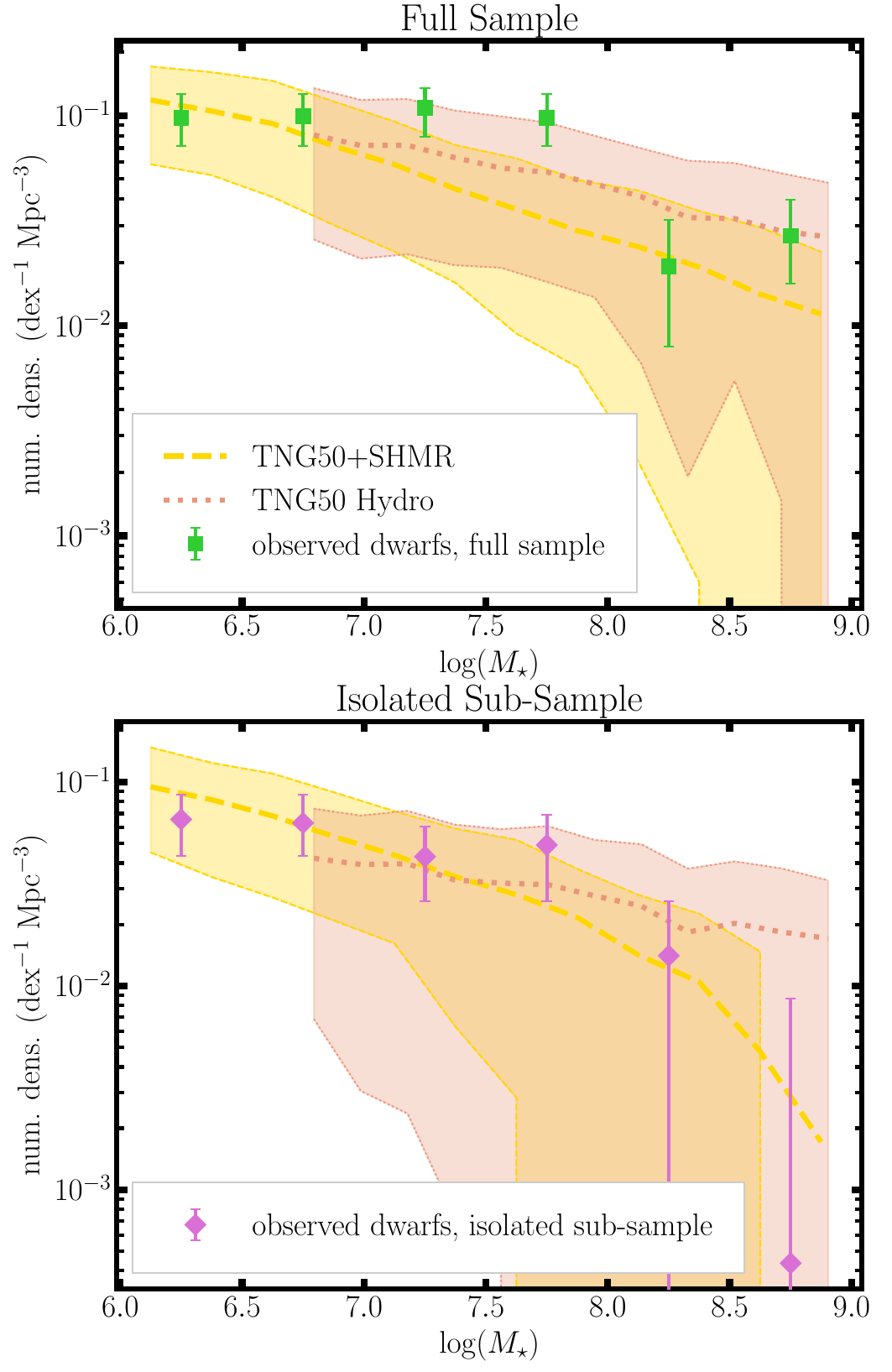}
\caption{A comparison between the observed dwarf SMFs and those predicted from the TNG50 simulation project. The simulations have been mock observed incorporating the observational footprint (Figure \ref{fig:survey_area}), detection pipeline sensitivity (Figure \ref{fig:completeness}), and distance measurement uncertainties. The top panel shows the entire dwarf population while the bottom panel shows the isolated sub-sample using the fiducial definition of `isolation' from \S\ref{sec:isolation}. SMFs from TNG50 are shown using both the hydrodynamic-based predicted stellar masses from TNG50 and stellar masses from the stellar-to-halo-mass relation from \citet{danieli2023}. The shaded bands show the $1\sigma$ spread in simulation predictions. Due to the relatively small area footprint of ELVES-Field, there is clearly significant stochasticity in the predicted dwarf abundance, and the observed abundance fits comfortably within the predicted scatter.  }
\label{fig:tng50}
\end{figure}

Figure \ref{fig:tng50} shows the comparison between the predicted dwarf abundance from TNG50 and the abundance found in ELVES-Field. We separately show the full dwarf sample (top) and the isolated field sub-sample (bottom) using the fiducial definition of `isolation' from \S\ref{sec:isolation}, applied to both the observations and simulations. The observed SMFs are the same as in Figure \ref{fig:smf}. We show both the predicted dwarf SMFs using the hydrodynamic results for stellar masses from TNG50 and using the SHMR from \citet{danieli2023}. The shaded bands show the $1\sigma$ spread in simulation results. It is clear in both panels that there is significant variation in the predicted dwarf abundance depending on the vantage point and orientation taken. This is sensible as ELVES-Field only covers $\lesssim 3000$ sq. deg., and there will be significant stochasticity in the number of dwarfs that fall into the footprint depending on where the footprint falls compared to the local cosmic structure (e.g. filaments or sheets vs. voids). With that said, the observed abundance of dwarfs (both total and isolated) in ELVES-Field falls comfortably within the expectations from the TNG50 simulation.  We note that the spread in the TNG50 simulation results does not seem to span high enough to include the lowest mass bin GAMA point from Figure \ref{fig:smf}.

Before moving on, we note that, ideally, the comparison with simulations would be done with simulations that are tuned to match the larger scale (on scales of tens of Mpc) environment of the MW. There are a number of existing simulation projects that attempt to do this on a variety of scales \citep[e.g.][]{gottloeber2010, sorce2016, libeskind2020, sawala2022, dolag2023}. Using these simulations instead of the environment-agnostic TNG50 could reduce the viewpoint-to-viewpoint scatter in field dwarf abundance that we find in TNG50.

\section{Summary}
\label{sec:conclusions}
In this paper, we present ELVES-Field, an extension of the original ELVES Survey \citep{carlsten2022}. We search for nearby Local Volume ($D<10$ Mpc) dwarf galaxies over $\sim 3,000$ square degrees of sky, focusing primarily on isolated dwarfs to complement ELVES, which focused on satellite dwarfs. In this survey, we measure the distance, or attempt to, for every detected candidate dwarf. 
%The primary science goals of the survey are 1) to infer the stellar mass function of field dwarfs (this work and 2) provide a first estimate of the quenched fraction of field dwarfs at low masses.  

The footprint is defined largely by where there is archival coverage of deep Subaru/HSC imaging (cf. Figure \ref{fig:survey_area}), which is critical in measuring the distance to detected candidate dwarfs. We detect dwarfs in Legacy Survey imaging \citep[LSDR10,][]{decals} using a specialized semi-automated pipeline focused on low surface brightness dwarfs (cf. Figure \ref{fig:det}). Using injection tests of artificial dwarfs, we establish our completeness to $r_e \sim 6$\arcs~ and $\mu_{\mathrm{eff}} < 27$ mag arcsec$^{-2}$ (cf. Figure \ref{fig:completeness}). Assuming LV dwarfs are uniformly distributed in space and follow the mass-size relation of \citet{carlsten2021a}, we infer a $50\%$ stellar mass completeness down to $M_\star \sim 10^6$ \msun\ (Figure \ref{fig:completeness2}). 

Once the candidate dwarfs are detected, we make an exhaustive effort to measure the distance to each to confirm or reject it as a genuine LV dwarf. We make use of tip of the red giant branch distances and redshifts, when available, but rely heavily on surface brightness fluctuation (SBF) measurements to constrain the distances to the majority of candidates. This measurement relies on the deep, high-quality HSC imaging that is available across the search footprint. With the distance measurements, we are able to conclusively confirm or reject a majority of the candidates (cf. Figure \ref{fig:psat}), however, a sizable fraction remain without conclusive distance constraints. For these, we develop a simple statistical model to estimate the probability that they are actual LV dwarfs.

For all the confirmed and candidate LV dwarfs, we provide S\'{e}rsic-based photometry. For the confirmed dwarfs, we also provide NUV photometry from archival GALEX observations where available. The GALEX data help determine the star-forming properties of the dwarfs, which is the focus of a companion paper. Using the group catalog of \citet{kourkchi2017}, we identify which LV dwarfs are likely isolated using various environmental criteria. We identify those dwarfs that are outside of a projected $1\times R_{\mathrm{vir}}$, $2\times R_{\mathrm{vir}}$, or $3\times R_{\mathrm{vir}}$ from any massive, $M_\star > 10^9$ \msun, group from \citet{kourkchi2017}. We use the $2\times R_{\mathrm{vir}}$ condition as our fiducial criterion for identifying isolated dwarfs. Additionally, we mimic the isolation criterion of \citet{geha2012} and identify which dwarfs are within a fixed projected 1.5 Mpc from any massive, $M_\star > 2.5 \times 10^{10}$ \msun, host galaxy. The lists of confirmed and candidate dwarfs we provide in Appendix \ref{app:photometry} have not had any environmental cut applied to them so that researchers can apply whichever cut is most appropriate to their use case.

The main goal of this paper is to present an overview of the catalog and provide the lists of detected dwarfs. We also present the stellar mass and luminosity functions of ELVES-Field dwarfs, both the entire sample and the isolated sub-sample (Figure \ref{fig:smf}). The two samples show similar slopes with the isolated sub-sample simply lower in normalization. The full sample shows fair agreement with the normalization of the GAMA \citep{driver2022} SMF at high dwarf masses ($10^8 < M_\star < 10^9$ \msun). In Figure \ref{fig:tng50}, we perform a simple comparison with the predicted SMF of dwarfs in the TNG50 hydrodynamic simulation \citep{tng50_1, tng50_2}. We forward model the TNG50 simulation incorporating our survey footprint, detection completeness, and distance measurement errors. We find that there is significant cosmic variance in the predicted dwarf SMF depending on vantage point in the simulation box but that the observed SMF fits comfortably in the predicted range. 

In a companion paper (Carlsten et al., accepted), we investigate the sizes and star-forming properties of the dwarf sample.

In the coming years, Vera Rubin Observatory's Legacy Survey of Space and Time will produce high quality imaging over a much larger area of the sky which will allow for a much larger sample of dwarfs to be curated using similar techniques as used here. In this way, ELVES-Field can be viewed as a test run for how the Rubin data can be used to actually produce a volume-limited sample of nearby dwarf galaxies. The larger sample will be critical in reducing cosmic variance in the stellar mass function and increasing the sample size of quenched, field dwarfs. Additionally, Euclid \citep[e.g.][]{marleau2025, hunt2025} and the Roman Space Telescope will allow for the procurement of a significant sample of field dwarfs using only tip of the red giant branch (i.e. no SBF), and it will be important to compare those results with an SBF-based sample.

\section*{Acknowledgements}
This work was supported by NSF AST-2006340. J.L. and J.E.G. gratefully acknowledge support from the NSF grant AST-2506292. J.L. acknowledges support from the Charlotte Elizabeth Procter Fellowship at Princeton University.

The Legacy Surveys consist of three individual and complementary projects: the Dark Energy Camera Legacy Survey (DECaLS; Proposal ID \#2014B-0404; PIs: David Schlegel and Arjun Dey), the Beijing-Arizona Sky Survey (BASS; NOAO Prop. ID \#2015A-0801; PIs: Zhou Xu and Xiaohui Fan), and the Mayall z-band Legacy Survey (MzLS; Prop. ID \#2016A-0453; PI: Arjun Dey). DECaLS, BASS and MzLS together include data obtained, respectively, at the Blanco telescope, Cerro Tololo Inter-American Observatory, NSF’s NOIRLab; the Bok telescope, Steward Observatory, University of Arizona; and the Mayall telescope, Kitt Peak National Observatory, NOIRLab. The Legacy Surveys project is honored to be permitted to conduct astronomical research on Iolkam Du’ag (Kitt Peak), a mountain with particular significance to the Tohono O’odham Nation.

NOIRLab is operated by the Association of Universities for Research in Astronomy (AURA) under a cooperative agreement with the National Science Foundation.

This project used data obtained with the Dark Energy Camera (DECam), which was constructed by the Dark Energy Survey (DES) collaboration. Funding for the DES Projects has been provided by the U.S. Department of Energy, the U.S. National Science Foundation, the Ministry of Science and Education of Spain, the Science and Technology Facilities Council of the United Kingdom, the Higher Education Funding Council for England, the National Center for Supercomputing Applications at the University of Illinois at Urbana-Champaign, the Kavli Institute of Cosmological Physics at the University of Chicago, Center for Cosmology and Astro-Particle Physics at the Ohio State University, the Mitchell Institute for Fundamental Physics and Astronomy at Texas A\&M University, Financiadora de Estudos e Projetos, Fundacao Carlos Chagas Filho de Amparo, Financiadora de Estudos e Projetos, Fundacao Carlos Chagas Filho de Amparo a Pesquisa do Estado do Rio de Janeiro, Conselho Nacional de Desenvolvimento Cientifico e Tecnologico and the Ministerio da Ciencia, Tecnologia e Inovacao, the Deutsche Forschungsgemeinschaft and the Collaborating Institutions in the Dark Energy Survey. The Collaborating Institutions are Argonne National Laboratory, the University of California at Santa Cruz, the University of Cambridge, Centro de Investigaciones Energeticas, Medioambientales y Tecnologicas-Madrid, the University of Chicago, University College London, the DES-Brazil Consortium, the University of Edinburgh, the Eidgenossische Technische Hochschule (ETH) Zurich, Fermi National Accelerator Laboratory, the University of Illinois at Urbana-Champaign, the Institut de Ciencies de l’Espai (IEEC/CSIC), the Institut de Fisica d’Altes Energies, Lawrence Berkeley National Laboratory, the Ludwig Maximilians Universitat Munchen and the associated Excellence Cluster Universe, the University of Michigan, NSF’s NOIRLab, the University of Nottingham, the Ohio State University, the University of Pennsylvania, the University of Portsmouth, SLAC National Accelerator Laboratory, Stanford University, the University of Sussex, and Texas A\&M University.

BASS is a key project of the Telescope Access Program (TAP), which has been funded by the National Astronomical Observatories of China, the Chinese Academy of Sciences (the Strategic Priority Research Program “The Emergence of Cosmological Structures” Grant \# XDB09000000), and the Special Fund for Astronomy from the Ministry of Finance. The BASS is also supported by the External Cooperation Program of Chinese Academy of Sciences (Grant \# 114A11KYSB20160057), and Chinese National Natural Science Foundation (Grant \# 11433005).

The Legacy Survey team makes use of data products from the Near-Earth Object Wide-field Infrared Survey Explorer (NEOWISE), which is a project of the Jet Propulsion Laboratory/California Institute of Technology. NEOWISE is funded by the National Aeronautics and Space Administration.

The Legacy Surveys imaging of the DESI footprint is supported by the Director, Office of Science, Office of High Energy Physics of the U.S. Department of Energy under Contract No. DE-AC02-05CH1123, by the National Energy Research Scientific Computing Center, a DOE Office of Science User Facility under the same contract; and by the U.S. National Science Foundation, Division of Astronomical Sciences under Contract No. AST-0950945 to NOAO.

The Hyper Suprime-Cam (HSC) collaboration includes the astronomical communities of Japan and Taiwan, and Princeton University. The HSC instrumentation and software were developed by the National Astronomical Observatory of Japan (NAOJ), the Kavli Institute for the Physics and Mathematics of the Universe (Kavli IPMU), the University of Tokyo, the High Energy Accelerator Research Organization (KEK), the Academia Sinica Institute for Astronomy and Astrophysics in Taiwan (ASIAA), and Princeton University. Funding was contributed by the FIRST program from the Japanese Cabinet Office, the Ministry of Education, Culture, Sports, Science and Technology (MEXT), the Japan Society for the Promotion of Science (JSPS), Japan Science and Technology Agency (JST), the Toray Science Foundation, NAOJ, Kavli IPMU, KEK, ASIAA, and Princeton University. 

This paper makes use of software developed for Vera C. Rubin Observatory. We thank the Rubin Observatory for making their code available as free software at http://pipelines.lsst.io/.

This paper is based on data collected at the Subaru Telescope and retrieved from the HSC data archive system, which is operated by the Subaru Telescope and Astronomy Data Center (ADC) at NAOJ. Data analysis was in part carried out with the cooperation of Center for Computational Astrophysics (CfCA), NAOJ. We are honored and grateful for the opportunity of observing the Universe from Maunakea, which has the cultural, historical and natural significance in Hawaii.

\software{ \texttt{astropy} \citep{astropy_2013, astropy_2018, astropy_2022}, \texttt{photutils} \citep{bradley2020}, \texttt{sep} \citep{sep}, \texttt{imfit} \citep{imfit}, \texttt{SWarp} \citep{swarp}, \texttt{Scamp} \citep{scamp}, \texttt{SExtractor} \citep{sextractor}, \texttt{astrometry.net} \citep{astrometry_net}   } 

\bibliographystyle{aasjournal}
\bibliography{calib}

\appendix

\section{Dwarf Photometry and Tables}
\label{app:photometry}
In this appendix, we provide lists of dwarfs detected in ELVES-Field. Table \ref{tab:conf_photo} provides the main photometry results for the confirmed LV dwarfs. Unless the dwarf has a previous name (e.g. in the LVGC), we name the dwarfs according to their RA/Dec coordinates. The table indicates whether the dwarf was confirmed via TRGB, redshift, or SBF. We also include a column for the `confirmed weight' that denotes the probability that the dwarf is truly within 10 Mpc given its measured distance and uncertainty. For example, a confirmed LV dwarf with a measured distance of $10\pm1.5$ Mpc will have a confirmed weight of 0.5. The table also includes several flags for the SBF measurement. The first flag indicates whether the dwarf used the default mask and annulus choices outlined in \S\ref{sec:sbf} or whether these were manually tweaked. The second flag indicates whether, if the mask and/or annulus were tweaked, the choice was particularly ambiguous. As discussed in \S\ref{sec:sbf}, different masks and/or annuli choices could have a material impact on the SBF distance, and these cases should be treated with caution. There are few enough of these cases to have no significant impact on the main science results, however. The third SBF flag in Table \ref{tab:conf_photo} indicates whether the SBF measurement was manually failed due to the fluctuation power not appearing to come from genuine SBF. We discuss this visual sanity check step in \S\ref{sec:dist_results}. In these cases, we conservatively keep the dwarf in the candidate (i.e. not `confirmed') status from the perspective of SBF. The only way a dwarf in Table \ref{tab:conf_photo} has this flag set if it is confirmed via one of the other methods (e.g. redshift). For dwarfs where no SBF measurement was attempted, the table simply lists `N/A'. Table \ref{tab:conf_photo} also includes flags for the inferred environment of the dwarfs. The first three flags indicate whether the dwarf is within either a projected $1\times R_{\mathrm{vir}}$, $2\times R_{\mathrm{vir}}$, or $3\times R_{\mathrm{vir}}$ of a massive group, respectively. The fourth flag indicates whether the dwarf is within a fixed 1.5 projected Mpc from any massive galaxy with $M_\star > 2.5\times10^{10}$\msun. More details on these isolation flags can be found in \S\ref{sec:isolation}. The convention for the isolation flags is that \texttt{False} means the dwarf is \textit{not} near a massive companion (i.e. is isolated). 

\startlongtable
\begin{longrotatetable}
\movetabledown=10mm
\begin{deluxetable}{cccccccccccccc}
\tablecaption{Confirmed Dwarf Photometry\label{tab:conf_photo}}
\tablehead{
\colhead{Name} & \colhead{RA} & \colhead{DEC}  & \colhead{Source} & \colhead{Dist.}  & \colhead{Conf. Wt.} & \colhead{$m_g$}  & \colhead{$m_r$}       & \colhead{$\log(M_\star)$}   & \colhead{$r_e$}   & \colhead{vel.}   & \colhead{SBF Flags}   & \colhead{Iso. Flags}   & \colhead{Quenched}       \\ 
\colhead{} & \colhead{(deg)}  & \colhead{(deg)}   & \colhead{}   & \colhead{Mpc} & \colhead{}  & \colhead{(mag)}   & \colhead{(mag)}        & \colhead{}   & \colhead{\arcs}   & \colhead{km/s}   & \colhead{}   & \colhead{}   & \colhead{}  }  
\startdata
IC1613 & 16.1992 & 2.1178 & T & 0.76$\pm$0.04 & 1.0 & 11.19$\pm$0.08 & 10.8$\pm$0.08 & 7.31$\pm$0.05 & 177.48$\pm$11.47 & -232.91 & N/A & T-T-T-T & False\\ 
UGC00685 & 16.8435 & 16.6845 & T & 4.81$\pm$0.24 & 1.0 & 13.73$\pm$0.08 & 13.34$\pm$0.08 & 7.85$\pm$0.05 & 26.63$\pm$1.72 & 151.04 & F-F-F & F-F-F-F & False\\ 
UGC00695 & 16.9436 & 1.0641 & R & 8.36$\pm$2.0 & 0.793 & 14.53$\pm$0.08 & 14.18$\pm$0.08 & 7.91$\pm$0.2 & 18.99$\pm$1.23 & 628.46 & F-F-F & F-T-T-F & False\\ 
UGC01056 & 22.1983 & 16.6892 & R & 8.86$\pm$2.0 & 0.716 & 14.71$\pm$0.08 & 14.28$\pm$0.08 & 8.01$\pm$0.19 & 14.97$\pm$0.97 & 592.59 & T-F-F & F-T-T-F & False\\ 
UGC01085 & 22.8284 & 7.7878 & R & 9.07$\pm$2.0 & 0.679 & 15.5$\pm$0.08 & 15.16$\pm$0.08 & 7.62$\pm$0.18 & 14.47$\pm$0.94 & 652.71 & T-F-F & F-F-F-F & False\\ 
dw0020p0837 & 5.1719 & 8.617 & R & 9.45$\pm$2.0 & 0.609 & 16.47$\pm$0.08 & 16.28$\pm$0.08 & 7.11$\pm$0.18 & 11.61$\pm$0.75 & 692.8 & F-F-F & F-F-F-F & False\\ 
PiscesA & 3.6917 & 10.8132 & T & 5.65$\pm$0.28 & 1.0 & 17.22$\pm$0.09 & 17.05$\pm$0.09 & 6.42$\pm$0.05 & 9.69$\pm$0.69 & 234.89 & F-F-F & F-F-F-F & False\\ 
dw0112p0129 & 18.0302 & 1.4843 & S & 13.21$\pm$1.95 & 0.05 & 19.49$\pm$0.19 & 19.02$\pm$0.2 & 6.65$\pm$0.15 & 6.21$\pm$1.07 &  & F-F-F & T-T-T-F & True\\ 
dw0136p1628 & 24.0842 & 16.4701 & S & 8.44$\pm$0.88 & 0.961 & 17.73$\pm$0.11 & 17.17$\pm$0.12 & 7.06$\pm$0.09 & 7.54$\pm$0.66 &  & T-F-F & T-T-T-F & True\\ 
dw0111p0227 & 17.9666 & 2.4628 & S & 9.25$\pm$1.49 & 0.691 & 19.52$\pm$0.19 & 19.29$\pm$0.21 & 6.02$\pm$0.16 & 5.45$\pm$0.95 &  & F-F-F & F-F-T-F & False\\ 
UGC01999 & 37.9691 & 19.1531 & R & 13.71$\pm$2.0 & 0.032 & 14.36$\pm$0.08 & 13.98$\pm$0.08 & 8.43$\pm$0.12 & 28.13$\pm$1.82 & 972.58 & T-F-F & F-F-F-F & False\\ 
UGC02168 & 40.2525 & 17.4327 & R & 11.83$\pm$2.0 & 0.181 & 15.45$\pm$0.08 & 15.16$\pm$0.08 & 7.79$\pm$0.14 & 20.12$\pm$1.3 & 789.04 & F-F-F & F-F-F-F & False\\ 
dw0243m0015 & 40.7935 & -0.2623 & R & 8.86$\pm$2.0 & 0.715 & 16.09$\pm$0.08 & 15.63$\pm$0.08 & 7.54$\pm$0.19 & 14.78$\pm$0.96 & 630.66 & T-F-F & T-T-T-T & False\\ 
dw0236m0001 & 39.1771 & -0.0171 & R & 13.79$\pm$2.0 & 0.029 & 16.11$\pm$0.08 & 15.87$\pm$0.08 & 7.61$\pm$0.12 & 16.06$\pm$1.04 & 1112.26 & F-F-F & T-T-T-T & False\\ 
dw0242p0000 & 40.5015 & 0.0146 & S & 11.64$\pm$1.51 & 0.139 & 16.12$\pm$0.08 & 15.54$\pm$0.08 & 7.92$\pm$0.11 & 9.09$\pm$0.59 & 1115.2 & T-F-F & T-T-T-T & True\\ 
dw0445p0344 & 71.4943 & 3.7474 & S & 5.46$\pm$0.54 & 1.0 & 17.21$\pm$0.09 & 16.82$\pm$0.09 & 6.67$\pm$0.09 & 11.62$\pm$0.82 &  & T-F-F & F-F-F-F & False\\ 
dw0249m0239 & 42.3434 & -2.6546 & R & 13.59$\pm$2.0 & 0.036 & 15.05$\pm$0.08 & 14.75$\pm$0.08 & 8.07$\pm$0.12 & 36.64$\pm$2.37 & 1093.49 & F-F-T & F-T-T-T & False\\ 
\enddata
\tablecomments{The main photometric results for the ELVES-Field Survey. Includes only dwarfs with conclusive distance confirmation. The `Source' column denotes what kind of distance measurement was used to confirm the dwarf as being in the LV: T=TRGB, R=Redshift, and S=SBF. The `Conf. Wt.' column denotes the probability the dwarf is within 10 Mpc based on its measured distance and uncertainty. The `SBF Flags' column denotes three separate flags regarding the SBF: 1) whether the dwarf used default mask/annulus parameters in the SBF measurement, 2) whether the resulting mask/annulus choices were particularly ambiguous, and 3) whether the dwarf failed the visual SBF check. More details on these flags is given in the text. Dwarfs for which no SBF measurement was attempted are indicated by a `N/A'. Finally, the `Iso. Flags' column denotes the environment of the dwarf. The first three flags indicate whether the dwarf is within either a projected $1\times R_{\mathrm{vir}}$, $2\times R_{\mathrm{vir}}$, or $3\times R_{\mathrm{vir}}$ of a massive group, respectively. The fourth flag indicates whether the dwarf is within 1.5 projected Mpc from any massive galaxy with $M_\star > 2.5\times10^{10}$\msun. The full table will be published online or will be provided upon request from the authors.}
\end{deluxetable}
\end{longrotatetable}

Table \ref{tab:cand_photo} provides the photometry for the candidate LV dwarfs (i.e. those without conclusive distance measurements). If the candidate dwarf has a distance lower bound set from the SBF measurement, that is listed in the table. Additionally, the estimated probabilities that the candidates are genuine LV dwarfs from the model described in \S\ref{sec:maybes} are also given.

\startlongtable
\begin{longrotatetable}
\movetabledown=10mm
\begin{deluxetable}{ccccccccccc}
\tablecaption{Candidate Dwarf Photometry\label{tab:cand_photo}}
\tablehead{
\colhead{Name} & \colhead{RA} & \colhead{DEC}  & \colhead{Dist. Low Bound} & \colhead{$P_{\mathrm{LV}}$} & \colhead{$m_g$}  & \colhead{$m_r$}        & \colhead{$r_e$}    & \colhead{SBF Flags}   & \colhead{Iso. Flags}   & \colhead{Quenched}       \\ 
\colhead{} & \colhead{(deg)}  & \colhead{(deg)}  & \colhead{Mpc} & \colhead{}  & \colhead{(mag)}   & \colhead{(mag)}      & \colhead{\arcs}   & \colhead{}   & \colhead{}   & \colhead{}  }  
\startdata
dw0122p0508 & 20.6364 & 5.138 & 9.75 & 0.0031 & 16.42$\pm$0.08 & 15.75$\pm$0.08 & 10.04$\pm$0.65 & F-F-F & F-F-F-F & False\\ 
dw0117p0958 & 19.4786 & 9.972 & 9.62 & 0.0032 & 17.39$\pm$0.09 & 16.76$\pm$0.1 & 6.39$\pm$0.49 & T-F-F & F-F-F-F & False\\ 
dw0157m0716 & 29.4745 & -7.2709 & 7.69 & 0.014 & 17.58$\pm$0.1 & 17.16$\pm$0.11 & 8.32$\pm$0.69 & T-F-F & F-F-F-F & False\\ 
dw0119p1038 & 19.8688 & 10.6429 & 6.62 & 0.0179 & 17.61$\pm$0.1 & 17.19$\pm$0.11 & 9.71$\pm$0.81 & T-F-T & F-F-F-F & False\\ 
dw0030p1002 & 7.6568 & 10.0407 & 6.64 & 0.0168 & 17.69$\pm$0.1 & 17.12$\pm$0.11 & 5.57$\pm$0.48 & T-F-F & F-F-F-F & False\\ 
dw0043p0150 & 10.8617 & 1.8476 & 9.66 & 0.0022 & 17.73$\pm$0.11 & 17.43$\pm$0.12 & 14.15$\pm$1.24 & F-F-F & F-F-F-F & False\\ 
dw0040p0245 & 10.2047 & 2.7578 & 8.83 & 0.007 & 17.77$\pm$0.11 & 17.43$\pm$0.12 & 4.91$\pm$0.44 & T-F-F & F-F-F-F & False\\ 
dw0030p1008 & 7.7187 & 10.1474 & 8.84 & 0.0062 & 17.9$\pm$0.11 & 17.2$\pm$0.12 & 6.23$\pm$0.59 & T-F-F & F-F-F-F & False\\ 
dw0121p0839 & 20.2759 & 8.6653 & 7.94 & 0.0083 & 18.11$\pm$0.12 & 17.6$\pm$0.13 & 5.27$\pm$0.54 & T-F-F & F-F-F-F & False\\ 
dw0116p0404 & 19.1631 & 4.0743 & 7.92 & 0.007 & 18.28$\pm$0.13 & 17.9$\pm$0.14 & 7.94$\pm$0.87 & T-F-F & F-F-F-F & False\\ 
dw0207p1819 & 31.7555 & 18.3181 & 9.9 & 0.0004 & 18.34$\pm$0.13 & 18.19$\pm$0.14 & 11.24$\pm$1.26 & T-F-F & F-F-F-F & False\\ 
dw0211p1039 & 32.9466 & 10.6556 & 7.73 & 0.0062 & 18.46$\pm$0.14 & 18.1$\pm$0.15 & 10.23$\pm$1.21 & T-F-F & F-F-F-F & False\\ 
dw0029m0026 & 7.3638 & -0.4352 & 8.62 & 0.0041 & 18.47$\pm$0.14 & 18.11$\pm$0.15 & 7.76$\pm$0.92 & T-F-F & F-F-F-F & False\\ 
dw0118p1019 & 19.6817 & 10.3275 & 8.61 & 0.0041 & 18.48$\pm$0.14 & 18.15$\pm$0.15 & 6.46$\pm$0.77 & T-F-F & F-F-F-F & False\\ 
dw0200p0248 & 30.0992 & 2.8068 & 8.6 & 0.0031 & 18.72$\pm$0.15 & 18.44$\pm$0.16 & 11.1$\pm$1.45 & T-F-F & F-F-F-F & False\\ 
dw0218m1204 & 34.5454 & -12.0697 & 7.61 & 0.0046 & 18.76$\pm$0.15 & 18.37$\pm$0.17 & 6.3$\pm$0.83 & T-F-F & F-F-F-T & False\\ 
dw0122p0840 & 20.7341 & 8.6736 & 7.77 & 0.0038 & 18.87$\pm$0.16 & 18.28$\pm$0.17 & 9.37$\pm$1.29 & F-F-F & F-F-F-F & False\\ 
\enddata
\tablecomments{Similar to Table \ref{tab:conf_photo} for the candidate dwarfs (i.e. dwarfs without conclusive distance results). If the candidate has a distance lower bound set from the SBF measurement, this is listed in the `Dist. Low Bound' column. The `$P_\mathrm{LV}$' column denotes the estimated probability that the dwarf is a true LV dwarf from the statistical model described in \S\ref{sec:maybes}. The `SBF Flags' and `Iso. Flags' columns are the same as in Table \ref{tab:conf_photo}. The full table will be published online or will be provided upon request from the authors.}
\end{deluxetable}
\end{longrotatetable}

Table \ref{tab:galex} provides the GALEX NUV photometry for the confirmed LV dwarfs that have archival GALEX coverage. In cases where the GALEX detection is $S/N < 2$, a $2\sigma$ lower limit to the NUV magnitude is given.

\begin{deluxetable*}{ccc}
\tablecaption{GALEX Photometry for confirmed dwarfs.\label{tab:galex}}
\tablehead{
\colhead{Name}  & \colhead{$m_\mathrm{NUV}$}  & \colhead{Lower Lim. Flag} \\ 
\colhead{}   & \colhead{(mag)}    & \colhead{} }  
\startdata
IC1613 &  & False\\ 
UGC00685 & 15.77$\pm$0.01 & False\\ 
UGC00695 & 16.31$\pm$0.01 & False\\ 
UGC01056 & 16.48$\pm$0.01 & False\\ 
UGC01085 & 17.24$\pm$0.04 & False\\ 
dw0020p0837 & 17.85$\pm$0.06 & False\\ 
PiscesA & 18.54$\pm$0.12 & False\\ 
dw0112p0129 & 23.54 & True\\ 
dw0136p1628 & 20.97$\pm$0.56 & False\\ 
dw0111p0227 & 21.23$\pm$0.15 & False\\ 
UGC01999 & 16.08$\pm$0.03 & False\\ 
UGC02168 & 16.63$\pm$0.04 & False\\ 
dw0243m0015 & 18.98$\pm$0.04 & False\\ 
dw0236m0001 & 17.79$\pm$0.05 & False\\ 
dw0242p0000 & 19.66$\pm$0.06 & False\\ 
dw0445p0344 &  & False\\ 
dw0249m0239 & 16.33$\pm$0.01 & False\\ 
\enddata
\tablecomments{The GALEX photometry for the distance-confirmed LV dwarfs that have archival coverage. For dwarfs with a detection with $S/N<2$, a $2\sigma$ lower limit to the UV magnitude is given, as indicated in the `Lower Lim. Flag' column. Empty entries indicate no GALEX coverage. The full version of the table will be published in the online journal or will be provided upon request to the authors.}
\end{deluxetable*}

Table \ref{tab:gaps} provides the locations and magnitude of detections in the LSDR10 footprint that did not have HSC coverage. These likely fell on chip gaps in the HSC data or immediately outside the F.O.V. of HSC pointings. These detections are not included in any analysis done here but are included for completeness.

\begin{deluxetable*}{ccc}
\tablecaption{Excluded detections not covered by HSC.\label{tab:gaps}}
\tablehead{
\colhead{RA}  & \colhead{DEC}  & \colhead{$m_g$} \\ 
\colhead{(deg)}   & \colhead{(deg)}    & \colhead{(mag)} }  
\startdata
27.9607 & -5.4961 & 14.96$\pm$0.08\\ 
27.9737 & -5.5103 & 15.06$\pm$0.08\\ 
12.9576 & 3.1055 & 15.45$\pm$0.08\\ 
19.7988 & 11.1212 & 16.95$\pm$0.08\\ 
21.0686 & -1.6713 & 16.97$\pm$0.08\\ 
18.7523 & 3.3055 & 17.1$\pm$0.08\\ 
0.5102 & -7.4627 & 17.19$\pm$0.09\\ 
20.1466 & 5.284 & 17.35$\pm$0.09\\ 
0.5835 & -7.6869 & 17.36$\pm$0.09\\ 
20.3926 & 5.7022 & 17.39$\pm$0.09\\ 
10.5789 & -1.4419 & 17.46$\pm$0.1\\ 
9.9598 & 3.1203 & 17.57$\pm$0.1\\ 
21.0541 & 5.1526 & 17.67$\pm$0.1\\ 
22.2249 & -1.6997 & 17.69$\pm$0.1\\ 
34.076 & 14.7844 & 17.84$\pm$0.11\\ 
21.2227 & -1.8448 & 17.98$\pm$0.12\\ 
21.7608 & 8.2669 & 18.1$\pm$0.12\\ 
\enddata
\tablecomments{The locations and magnitude of detections in the LSDR10 footprint that did not have coverage in HSC (e.g. fell in a chip gap or outside of the F.O.V.). These are not included in any analysis in the paper but are included here for completeness.}
\end{deluxetable*}

Table \ref{tab:rej} provides the locations and distance information for rejected candidates. We indicate whether the candidate was rejected based on TRGB, SBF, or redshift. If available, we provide the SBF distance lower bound and/or velocity of the candidate.

\begin{deluxetable*}{ccccc}
\tablecaption{Properties of rejected candidates.\label{tab:rej}}
\tablehead{
\colhead{RA}  & \colhead{DEC}  & \colhead{Rej. Source}  & \colhead{Dist. Lower Bound} & \colhead{vel.} \\ 
\colhead{(deg)}  & \colhead{(deg)}    & \colhead{}     & \colhead{(Mpc)}    & \colhead{(km/s)}  }  
\startdata
31.1309 & -6.199 & R &  & 1365.71 \\ 
 4.2003 & 12.3496 & R &  & 1142.18 \\ 
 8.3419 & -1.1216 & R & 7.74 & 1994.84 \\ 
 23.3275 & 3.0708 & R & 4.3 & 1720.94 \\ 
 27.5443 & 2.3102 & R & 8.17 & 1699.82 \\ 
 20.0282 & -0.2052 & S & 13.3 & 1727.18 \\ 
 13.22 & 1.214 & S & 11.65 & 1782.3 \\ 
 28.8975 & 3.2907 & R & 7.33 & 1752.12 \\ 
 4.0727 & 12.415 & R & 8.98 & 1659.59 \\ 
 18.7783 & -1.2742 & S & 14.02 &  \\ 
 20.5199 & 5.791 & S & 12.12 &  \\ 
 34.0042 & -2.0173 & S & 11.81 &  \\ 
 21.5205 & 0.3162 & R & 8.15 & 1927.19 \\ 
 17.2832 & 1.121 & R & 9.99 & 1169.28 \\ 
 9.8286 & 2.9872 & S & 11.79 &  \\ 
 17.2918 & 1.2911 & S & 10.28 & 1099.01 \\ 
 10.0109 & 3.0979 & S & 10.65 &  \\ 
 \enddata
\tablecomments{The properties of rejected candidates. `Rej. Source' indicates the distance source used in the rejection: T=TRGB, S=SBF, and R=redshift. If the candidate has a distance lower bound from an SBF measurement and/or a velocity, it is listed. The full version of the table will be published in the online journal or will be provided upon request to the authors.}
\end{deluxetable*}

\section{Tests of SBF Methodology}
\label{app:sbf_tests}
The SBF measurement methodology we use in ELVES-Field differs meaningfully from that used in the ELVES Survey \citep{carlsten2022}. Thus, we carefully tested it both using image simulations and by measuring the distance to dwarfs in the HSC footprint that have a known reference distance, either from TRGB, redshift, or a previously published SBF distance from ELVES. In the image simulations, we inject artificial dwarfs of typical brightness and size with SBF onto random fields in the HSC footprint. We then measure their SBF in the systematic way described in \S\ref{sec:sbf}. We find that we can recover the input SBF magnitude without bias and with a precision that is roughly the same as the estimated measurement uncertainty. 

For the test on real galaxies, we include two groups of test galaxies. The first is known nearby galaxies ($D\lesssim 12$ Mpc) from the Local Volume Galaxy Catalog \citep{karachentsev2013} and ELVES Survey that are in the HSC footprint. This is a subset of the reference sample discussed in \S\ref{sec:completeness} and shown in Figure \ref{fig:completeness}. The second is Sienna Galaxy Atlas \citep[SGA;][]{moustakas2023} dwarf galaxies with redshifts in the range $1500 < cz < 2200$ km/s. This redshift range is chosen to select galaxies that are nearby but still almost certainly outside of the LV (i.e. $D>10$ Mpc). In selecting these SGA galaxies, we additionally require them to pass the photometric cuts of Figure \ref{fig:photo_cuts} to select only dwarf galaxies. This SGA sample is used as a reference sample of `background' dwarfs to demonstrate that the SBF measurement process can correctly reject them as LV members.

\begin{figure*}
\includegraphics[width=\textwidth]{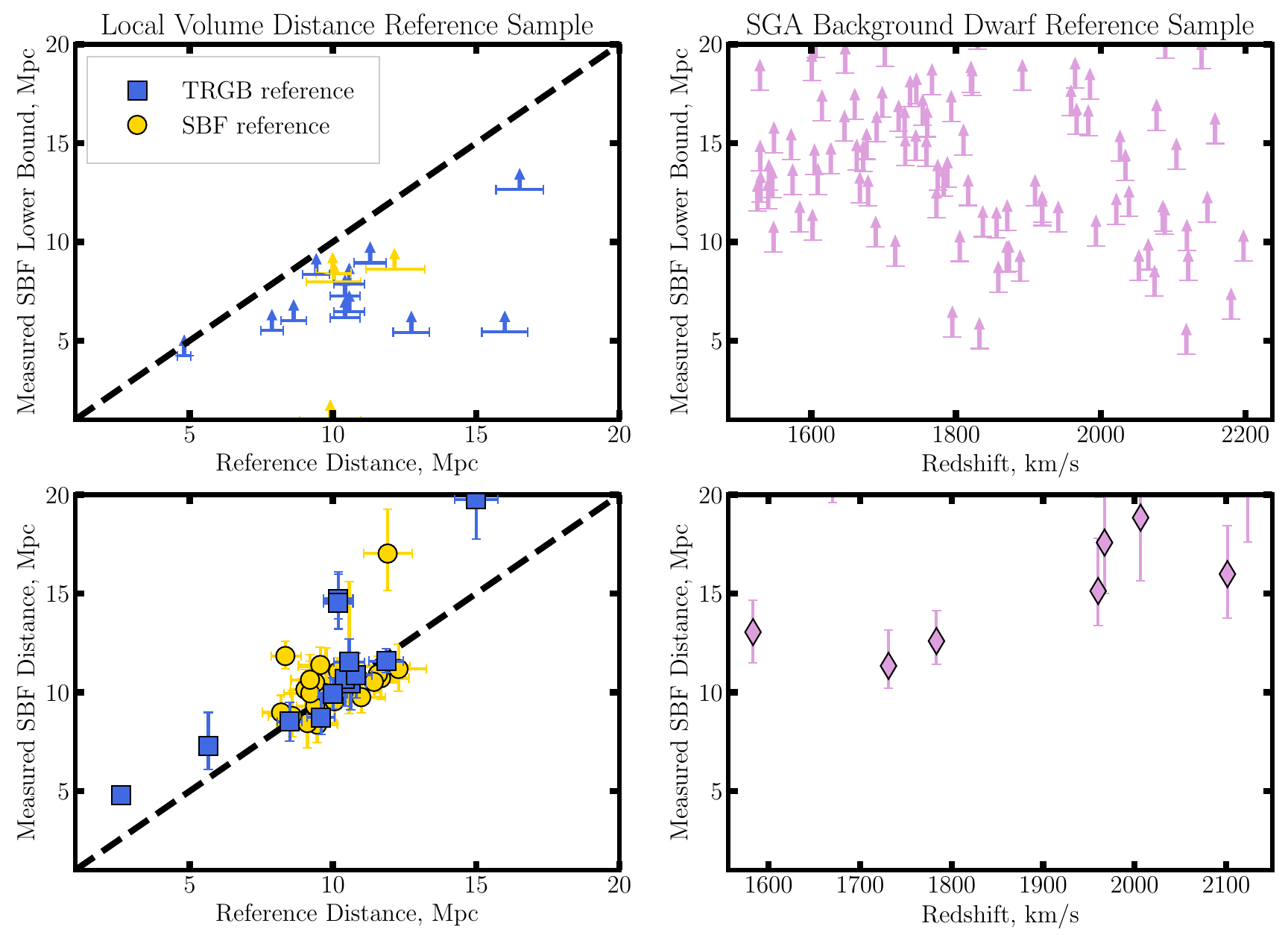}
\caption{To test the SBF measurement pipeline used in this work, we apply it to a sample of test galaxies with published reference distances. The left panels show a sample of known nearby galaxies ($D\lesssim 12$ Mpc) with either TRGB or previous SBF reference distances. The right panels show a sample of `background' galaxies from the SGA \citep{moustakas2023} whose redshifts indicate they are somewhat local but almost certainly beyond the LV ($D>10$ Mpc). The galaxies are further split based on whether they fail (top panels) or pass (bottom panels) the SBF `trustworthiness' flag defined in \S\ref{sec:sbf}. Galaxies that fail this flag likely have additional sources of fluctuation power that is biasing the SBF measurement too high. For these galaxies, we still use the SBF measurement to define a $2\sigma$ distance lower bound, as indicated by the upward arrows in the top panels. The top left panel shows that these distance lower bounds are valid compared to the known, reference distances. The top right panel shows that the majority of SGA galaxies fail the `trustworthiness' flag but the SBF measurements still constrain most to be beyond the LV. The bottom left panel shows good agreement between the measured SBF distance in the current work with the reference distance, except for a handful of dwarfs above the $1:1$ line whose distances in the automated SBF pipeline are overestimated. As discussed in the text, these cases suffer from clear overmasking using the automated mask which causes erroneously low SBF signal. These cases would be manually adjusted in the real, semi-automated measurement process.  Finally, the bottom right panel shows that the few SGA galaxies that pass the flag all have SBF distances $>10$ Mpc. }
\label{fig:sbf_test}
\end{figure*}

For each of these galaxies, we analyze them with the SBF measurement pipeline described in \S\ref{sec:sbf}. For this test, we do not manually adjust any of the masks or annuli used and rely only on the algorithmically determined masks and annuli. The results are shown in Figure \ref{fig:sbf_test}. In the end, this test includes 50 known LV members and 105 SGA `background' galaxies. The left panels show the LV galaxies and the right show the SGA galaxies. The left panels compare the measured SBF distances using the current pipeline to the reference distances published in the literature. The LV galaxies are split based on whether their reference distance is a TRGB or SBF distance from the ELVES Survey. The test galaxies are also split based on whether they have the SBF `trustworthiness' flag set, as defined in \S\ref{sec:sbf}. The galaxies in the top panel fail this flag and we show only the $2\sigma$ lower bound distances from the SBF pipeline. The galaxies all fall below the $1:1$ line, indicating that the distance lower bound set by the SBF pipeline is appropriate. The top right panel shows that the majority of the SGA sample is indeed shown to be in the background. Note that the majority of SGA galaxies fail the `trustworthiness' flag since most are blue, irregular galaxies. 

The galaxies in the bottom panels pass the `trustworthiness' flag, and we show the actual measured SBF distance. For the most part, there is good consistency between the measured SBF distance and the reference distance for the LV sample. The exception is a handful of galaxies above the $1:1$ line where the distance measured here is overestimated compared to the reference distance. Each of these cases suffers from obvious overmasking using the pipeline mask which causes erroneously low SBF signal. These cases would be fixed in the real measurement process as the masks would be manually adjusted. Even without incorporating the manual fixes, the SBF distances from the current pipeline are not greatly biased with respect to the reference distances. The median residual between the current SBF distance and the reference distance is only $2.5\%$, and the mean residual is $10\%$ due to the handful of outliers. In general, if the SBF distances were systematically biased high, the inferred physical sizes of the dwarfs would also be biased too large. In a companion paper, we present the mass-size relation of isolated ELVES-Field dwarfs and find them to actually be offset to smaller sizes than ELVES Satellites. This finding that field dwarfs are smaller than satellites at the same stellar mass has also recently been found by \citet{asali2025}.

The bottom right panel shows the few SGA galaxies with trustworthy SBF distances are all $>10$ Mpc.

\end{CJK*}
\end{document}